\UseRawInputEncoding
\documentclass[twocolumn, trackchanges]{aastex63}
\usepackage[utf8]{inputenc}
\usepackage{appendix}
\usepackage[normalem]{ulem}
\usepackage{soul}
\usepackage{cancel}
\usepackage{hyperref}

\usepackage{outlines}
\usepackage{multirow}
\usepackage{booktabs}

\shorttitle{Line-by-line velocity measurements}
\shortauthors{Artigau et al.}

\begin{document}

\title{Line-by-line velocity measurements, an outlier-resistant method for precision velocimetry}
\correspondingauthor{\'Etienne Artigau}
\email{etienne.artigau@umontreal.ca}

\author[0000-0003-3506-5667]{\'Etienne Artigau}
\affiliation{Universit\'e de Montr\'eal, D\'epartement de Physique, IREX, Montr\'eal, QC H3C 3J7, Canada}
\affiliation{Observatoire du Mont-M\'egantic, Universit\'e de Montr\'eal, Montr\'eal, QC H3C 3J7, Canada}
\author[0000-0001-9291-5555]{Charles Cadieux}
\affiliation{Universit\'e de Montr\'eal, D\'epartement de Physique, IREX, Montr\'eal, QC H3C 3J7, Canada}
\author[0000-0003-4166-4121]{Neil J. Cook}
\affiliation{Universit\'e de Montr\'eal, D\'epartement de Physique, IREX, Montr\'eal, QC H3C 3J7, Canada}
\author[0000-0001-5485-4675]{Ren\'e Doyon}
\affiliation{Universit\'e de Montr\'eal, D\'epartement de Physique, IREX, Montr\'eal, QC H3C 3J7, Canada}
\affiliation{Observatoire du Mont-M\'egantic, Universit\'e de Montr\'eal, Montr\'eal, QC H3C 3J7, Canada}
\author[0000-0002-5922-8267]{Thomas Vandal}
\affiliation{Universit\'e de Montr\'eal, D\'epartement de Physique, IREX, Montr\'eal, QC H3C 3J7, Canada}
\author[0000-0001-5541-2887]{Jean-Fran\c cois Donati}
\affiliation{Universit\'e de Toulouse, CNRS, IRAP, 14 Avenue Belin, 31400 Toulouse, France}
\author[0000-0002-2842-3924]{Claire Moutou}
\affiliation{Universit\'e de Toulouse, CNRS, IRAP, 14 Avenue Belin, 31400 Toulouse, France}
\author[0000-0001-5099-7978]{Xavier Delfosse}
\affiliation{Universit\'e Grenoble Alpes, CNRS, IPAG, 38000 Grenoble, France}
\author[0000-0002-1436-7351]{Pascal Fouqu\'e}
\affiliation{Universit\'e de Toulouse, CNRS, IRAP, 14 Avenue Belin, 31400 Toulouse, France}
\affiliation{Canada-France-Hawaii Telescope, CNRS, Kamuela, HI 96743, USA}

\author[0000-0002-5084-168X]{Eder Martioli}
\affiliation{Laborat\'orio Nacional de Astrof\'isica, Rua Estados Unidos 154, Itajub\'a, MG 37504-364, Brazil}\affiliation{Sorbonne Universit\'e, CNRS, UMR 7095, Institut d'Astrophysique de Paris, 98 bis bd Arago, 75014 Paris, France}
\author[0000-0002-7613-393X]{Fran\c cois Bouchy}
\affiliation{D\'epartement d'astronomie, Universit\'e de Gen\`eve, Chemin des Maillettes 51, CH-1290 Versoix, Switzerland}

\author[0000-0002-6013-4655]{Jasmine Parsons}
\affiliation{McGill University, Department of Physics \& McGill Space Institute,
3600 Rue University,
Montr\'eal, QC H3A 2T8, Canada}
\affiliation{Universit\'e de Montr\'eal, D\'epartement de Physique, IREX, Montr\'eal, QC H3C 3J7, Canada}

\author[0000-0003-2471-1299]{Andres Carmona}
\affiliation{Universit\'e Grenoble Alpes, CNRS, IPAG, 38000 Grenoble, France}

\author[0000-0002-9332-2011]{Xavier Dumusque}
\affiliation{D\'epartement d'astronomie, Universit\'e de Gen\`eve, Chemin des Maillettes 51, CH-1290 Versoix, Switzerland}

\author[0000-0002-8462-515X ]{Nicola Astudillo-Defru}
\affiliation{Departamento de Matem\'atica y F\'isica Aplicadas, Universidad Cat\'olica de la Sant\'isima Concepci\'on, Alonso de Rivera 2850, Concepci\'on, Chile}

\author[0000-0001-9003-8894]{Xavier Bonfils}
\affiliation{Universit\'e Grenoble Alpes, CNRS, IPAG, 38000 Grenoble, France}

\author[0000-0002-5407-3905]{Lucille Mignon}
\affiliation{Universit\'e Grenoble Alpes, CNRS, IPAG, 38000 Grenoble, France}

\begin{abstract}
We present a new algorithm for precision radial velocity (pRV) measurements, a line-by-line (LBL) approach designed to handle outlying spectral information in a simple but efficient manner.  The effectiveness of the LBL method is demonstrated on two datasets, one obtained with SPIRou on Barnard's star, and the other with HARPS on Proxima Centauri. In the near-infrared, the LBL provides a framework for m/s-level accuracy in pRV measurements despite the challenges associated with telluric absorption and sky emission lines. We confirm with SPIRou measurements spanning 2.7\,years that the candidate super-Earth on a \hbox{233-day} orbit around Barnard's star is an artifact due to a combination of time-sampling and activity. The LBL analysis of the Proxima Centauri {HARPS} post-upgrade data alone easily recovers the  Proxima\,b signal and also provides a 2-$\sigma$ detection of the recently confirmed 5-day Proxima\,d planet, but argues against the presence of the candidate Proxima\,c with a period of 1900\,days. We provide evidence that the Proxima\,c signal is associated with small, unaccounted systematic effects affecting the \texttt{HARPS-TERRA} template matching RV extraction method for long period signals. Finally, the LBL framework provides a very effective activity indicator, akin to the full width at half maximum derived from  the cross-correlation function, from which we infer a rotation period of $92.1^{+4.2}_{-3.5}$ days for Proxima.

\end{abstract}
\keywords{near-infrared velocimetry, Gliese 699, Proxima Centauri}

\section{Introduction}

Within the last three decades, the search for planets orbiting stars other than our Sun has moved from a speculative endeavour to one of the most active fields in astronomy. The first successful technique to uncover exoplanets around stars was the radial velocity method \citep{lindegren_fundamental_2003}, where one measures the reflex motion of a star as it is orbited by a low-mass companion. While that technique has been superseded in the raw number of discoveries by the transit method, 
radial velocity measurements remain central in exoplanet studies for providing mass measurements. Since the watershed discovery of the planet orbiting 51\,Pegasi, state-of-the-art accuracy achieved by this technique has been lowered from $\sim15$\,m/s \citep{mayor_jupiter-mass_1995} to better than 20\,cm/s \citep{faria_candidate_2022}.

While early precision radial velocity (pRV) efforts focused on Sun-like stars, M dwarfs are now the topic of active research. Having a relatively low mass (0.078 to $\sim$0.6\,M$_\odot$ \citealt{baraffe_new_2015}) and small radius (\hbox{0.1--0.6\,R$_\odot$}), a given planet will impart a larger RV signal and a much deeper transit than if it were orbiting a Sun-like star. When focusing on planets located in the habitable zone, this effect is further enhanced: habitable zones being more compact for M dwarfs than for Sun-like stars. This allows for m/s accuracy measurements to detect Earth-mass planets in the habitable zone of their parent stars (e.g., \citealt{bonfils_harps_2013, delfosse_harps_2013, anglada-escude_terrestrial_2016}), a feat yet to be achieved for such a planet around a Sun-like star, for which the RV signal would be at the level of 10\,cm/s.

The extension of pRV to late-type stars, combined with the interest of transit spectroscopy which displays numerous  atmospheric spectroscopic features in the near-infrared (e.g., He at 1.083\,$\mu$m to the CO bands starting at 2.29\,$\mu$m) has led to the development of pRV spectrographs covering the $YJHK$ bandpasses \citep{artigau_spirou_2012, quirrenbach_carmenes_2018, kotani_infrared_2014, claudi_giarpstng:_2017, mahadevan_habitable-zone_2012}. 

Precision spectroscopy in the near-infrared (nIR) is significantly more challenging than in the optical domain for a number of reasons.  Firstly, telluric absorption bands are much more numerous in the $YJHK$ domain, adding an imprint at near-zero velocity on all spectroscopic observations. Left uncorrected, biases due to telluric absorption would completely overwhelm planetary signals. A number of approaches have been proposed to correct telluric absorption in the near-infrared using both model-based (e.g., \citealt{smette_molecfit_2015, gullikson_telfit_2014}), data-based methods \citep{artigau_telluric-line_2014, bedell_wobble_2019, cretignier_yarara_2021} or the observation of hot stars shortly before or after the science observation \citep{vacca_method_2003}. The near-infrared night sky also displays strong emission features due mostly to OH airglow \citep{rousselot_night-sky_2000}.

At the typical resolution of pRV spectrographs ($\frac{\lambda}{\Delta\lambda}\sim50\,000$--100\,000), emission lines occupy only a small fraction of resolution elements (typically percent-level), but are strong enough to bias the RV measurement especially for stars near the ecliptic pole with small barycentric radial velocity (BERV).
Other effects that are specific to near-infrared spectrographs involve the significantly worse detector cosmetics compared to optical spectrographs. The state-of-the-art detectors used in near-infrared pRV spectrographs are Hawaii-4RG or Hawaii-2RG mosaics, which have dark current and cosmetics (e.g., bad pixel clusters, hot pixel clusters, dead columns, persistence) that lead to fixed structures in the extracted spectra which may bias pRV observations \citep{artigau_h4rg_2018}. Last but not least, nIR pRV requires a cryogenic spectrograph with thermal control at a level of 1\,mK, which in itself has been a major engineering challenge for all instruments with similar science goals (e.g., \citealt{donati_spirou_2020, Stefansson_versatile_2016, mirabet_carmenes_2014}).

This paper presents a new method for radial velocity measurements that is tailored to spectra having spurious excursions in flux from unaccounted effects: the line-by-line (LBL; see Appendix~\ref{code_accessibility} for code availability) pRV method. The technique has been proposed by \citet{dumusque_measuring_2018} and further explored by \citet{cretignier_measuring_2020} in the context of optical pRV observations and mostly tailored toward filtering stellar activity in Solar-type stars. Here, the concept is expanded to include the full radial-velocity content of molecular bands of M dwarfs and account for the needs of nIR pRV measurements where resilience against outliers is key. Section~\ref{prvmethods} provides the mathematical background pertaining to the discussion as well as a brief overview of other pRV numerical techniques. Section~\ref{demonstration_dataset} presents the datasets used for the demonstration of the LBL method. The mathematical framework of the LBL algorithm is described in Section~\ref{lbl_method}.  
 Section~\ref{demonstrations} presents a demonstration of the LBL technique based on SPIRou and HARPS datasets on Barnard's star \citep{barnard_small_1918} and Proxima Centauri \citep{innes_faint_1915}, respectively. We conclude with Section~\ref{conclusions}.

\section{pRV measurement methods\label{prvmethods}}

\subsection{Bouchy et al. 2001 mathematical framework}
\label{B01}
\citealt{bouchy_fundamental_2001} (B01 hereafter) presents a framework that can be used to reduce the problem of pRV measurements in noisy data to simple arithmetic; this works expands on a formalism first introduced by \citet{connes_demonstration_1996}. For pixel $i$, if one subtracts a spectrum $A(i)$ with infinitesimal RV shift $\delta V$ (or, expressed as a fractional wavelength change, $\frac{\delta V}{c} = \frac{\delta \lambda}{\lambda}$) from a zero velocity template $A_0(i)$, the residual ought to be the derivative of the template times the velocity change. More explicitly, it corresponds to the derivative of the template with respect to the wavelength grid  expressed as a velocity.  In this context, Equation (2) in B01 is:
\begin{equation}
A(i) - A_0(i) = \frac{\partial A_0(i)}{\partial \lambda(i)} \delta \lambda(i)
\label{eq:b01}
\end{equation}
which can be expressed as a velocity change

\begin{equation}
\frac{\delta V(i)}{c} =  \frac{A(i) - A_0(i)}{\lambda(i)\frac{\partial A_0(i)}{\partial\lambda(i)}}
\label{sum0}
\end{equation}

With an estimate of the local noise $\sigma(i)$ (\hbox{$[A(i)-A_0(i)]_{\rm RMS}$} in B01), which we obtain through a running standard deviation estimate of the $A(i) - A_0(i)$, one can determine the per-pixel RV accuracy $\sigma_V(i)$.  In our implementation, the noise estimate is computed with a running box the size of 500\,km/s (same size as the high-pass filtering width; see Section~\ref{high_pass_filter}) by taking half the difference between the 84$^{\rm th}$ and 16$^{\rm th}$ percentiles of residuals of $A(i)-A_0(i)$. This departs slightly from the B01 and \citet{dumusque_measuring_2018} implementation of the method as we do not explicitly assume that the noise is solely due to the combined contributions of Poisson and readout noise in the spectrum. An empirical determination of the noise relative to the template may be important when the detector readout-noise contribution is not straightforward to determine (e.g., spatially correlated readout-noise in IR arrays, integration-time dependent readout noise in up-the-ramp readout schemes)  or when modal noise for optical elements is a significant contribution \citep{blind_few-mode_2017}.  With the definition above for $\sigma(i)$, Equation~\ref{sum0} becomes:

\begin{equation}
\sigma_V(i) =  \frac{c\sigma(i)}{\lambda(i)\frac{\partial A_0(i)}{\partial\lambda(i)} }
\label{eq:noise}
\end{equation}

Considering that an optimal weighted sum is performed with weights proportional to the inverse of the square of uncertainties, one therefore has a mean velocity for the pixels of interests. That is:

\begin{equation}
\delta V = \frac{\sum  \delta V(i)  \sigma_V(i)^{-2} }{\sum \sigma_V(i)^{-2}}
\label{sum_rv}
\end{equation}
The  uncertainty of a weighted sum becomes:

\begin{equation}
\sigma_V = \frac{1}{\sqrt{\sum \sigma_V(i)^{-2}}}
\label{err_rv}
\end{equation}

The B01 formalism is, by construction, optimal in terms of signal-to-noise and provides a limit that cannot be surpassed in terms of RV accuracy for a given spectrum and noise.

A key point that will be used in the following sections is that summations in Equations \ref{sum_rv} and \ref{err_rv} can be performed over an interval in wavelength that is of arbitrary length. 


\subsection{The Cross-Correlation Function method}

One of the earliest methods to derive precise radial velocities from high-resolution spectroscopy is the cross-correlation function (CCF). The CCF method has its historical roots in pre-CCD spectroscopy \citep{baranne_coravel_1979}, where a mechanical mask was superposed onto a cross-dispersed spectrum and the total flux passing though holes corresponding to line positions was integrated with a single-pixel photoelectric sensor. By moving the mask mechanically, one could directly record the mean line profile. In modern spectrographs, the spectrum is cross-correlated with a comb of delta functions of varying strength matching known stellar lines  \citep{  pepe_harps:_2002, baranne_elodie_1996,lafarga_carmenes_2020}. One then derives a weighted mean line profile, the weight of each line scaling as its depth and local signal-to-noise ratio (SNR). The mean line SNR is far larger than that of the input spectrum and scales as the square root of the number of lines. The CCF provides a central velocity measurement and the higher moments of the line profiles inform on stellar activity (e.g., \citealt{santos_bisector_2001}).  One significant advantage of the CCF compared to other methods is that it  provides a velocity measurement with a single spectrum, assuming that one has a mask for the relevant spectral type. It also provides more complex information on line structures, such as the presence of double lines in a spectroscopic binary, that are not explicitly retrieved by other pRV measurement methods.

As detailed in Section~\ref{gains}, pRV in the nIR domain is strongly hampered by telluric absorption residuals and its time-varying impact on a given line profile. These effects  limit the usefulness of the CCF method at these wavelengths.

\subsection{Template matching}
While simple, robust and powerful,  the CCF technique does not fully exploit the radial velocity content \citep{anglada-escude_HARPS-TERRA_2012}. A number of groups have devised numerical methods where a high SNR template is adjusted to a given spectrum to measure its velocity. Given the proper weights, as described in Section~\ref{B01}, this method can fully exploit the  RV content of the observation. 

 Although the numerical details vary between implementations, template matching is now the most widely used method for pRV measurements of late-type stars with rich molecular bands. Virtually all recent measurements of low-mass planets around M dwarfs have employed this type of technique. The main template matching implementations widely used in the literature are \texttt{HARPS-TERRA} \citep{anglada-escude_HARPS-TERRA_2012}, \texttt{NAIRA} \citep{astudillo-defru_harps_2017}, \texttt{WOBBLE} \citep{bedell_wobble_2019} and \texttt{SERVAL} \citep{zechmeister_spectrum_2018}.

\section{Demonstration datasets\label{demonstration_dataset}}

In describing the LBL pRV measurement algorithms below, one needs sample datasets to perform the demonstration. As the initial motivation behind our implementation of the LBL method is to improve pRV measurements resilience to outliers in infrared observations, we used SPIRou data on Barnard's star, known to be stable within a few m/s. To compare the performances of the LBL method in a domain with  few telluric lines with a publicly available benchmark analysis, we also performed  LBL analysis using HARPS data on Proxima Centauri.

\subsection{SPIRou \label{spirou} data on Barnard's star}
Observations on Barnard's star were obtained with SPIRou as part of the SPIRou Legacy Survey \citep{donati_spirou_2020} between September 2018 and October 2021. The typical Barnard's star observation has a per-pixel SNR of 100, 150, 200 and 200 in $Y$, $J$, $H$ and $K$ photometric bandpasses respectively. 

\begin{figure*}[!htbp]
    \centering
    \includegraphics[width=\linewidth]{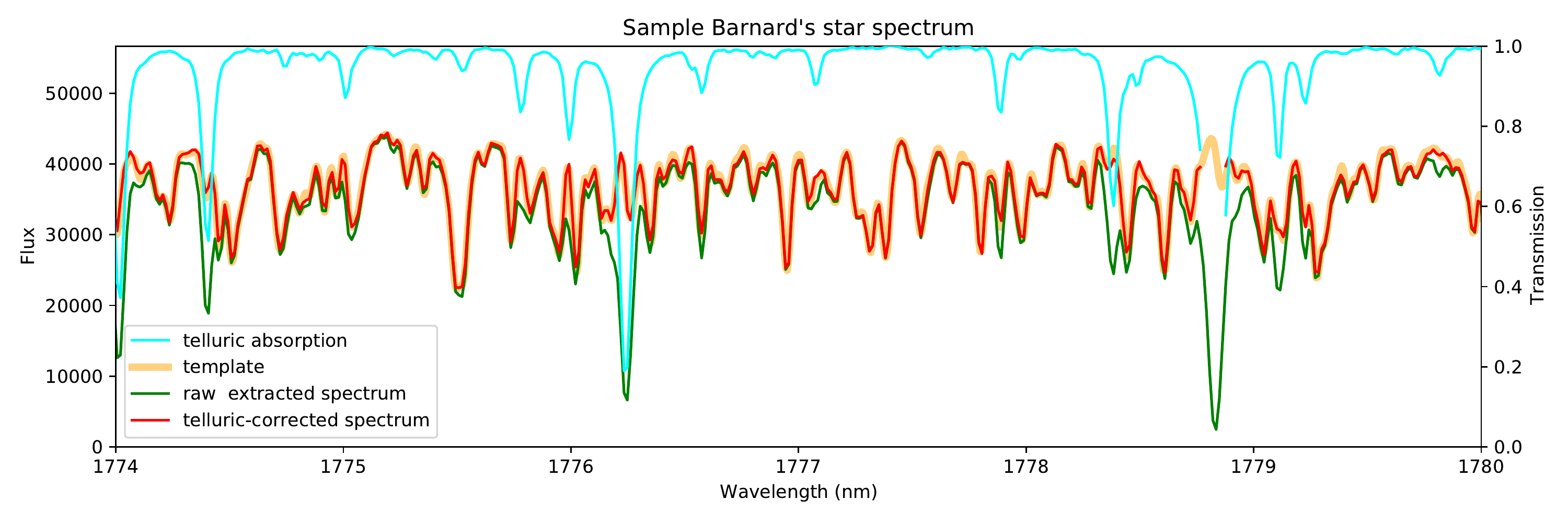}
    \caption{Sample spectrum of Barnard's star over a short domain wavelength within $H$ band. The telluric-corrected observation is over-plotted with the template registered to that star line-of-sight velocity accounting for the BERV at the time of observation.}
    \label{fig:gl699_spectrum}
\end{figure*}

The SPIRou data presented here was reduced with the version 0.7.194 of \texttt{APERO} (A PipelinE to Reduce Observations; Cook et al., in prep.). \texttt{APERO} performs all the steps from the raw 2D science frames to science-ready, wavelength calibrated data and the main data reduction steps are as follows. \texttt{APERO} starts with science frames that are constructed from the detector control software; these $4096\times4096$ frames represent the per-pixel accumulation rate of incoming flux (ramp images) and are corrected for non-linearity. \texttt{APERO} performs a set of quality checks on these frames to identify spurious pixels and filters correlated noise between amplifiers before the more standard extraction procedures. Frames are registered to a common position to account for slow bulk shifts/rotations of the science array, the 49 diffraction orders are identified and extracted optimally using the \citet{horne_optimal_1986} formalism while accounting for the shape of the SPIRou pupil slicer \citep{micheau_spirou_2018}. The product of the \texttt{APERO} extraction is $E2DS$ ($4088\times49$ pixels), an array consisting of all extract spectral orders. Once science data and calibrations have been extracted, wavelength calibration is performed similarly to the scheme described in \citet{hobson_spirou_2021}. As SPIRou is a spectropolarimeter with two science fibers (A and B) as well as a calibration (fiber C), one can extract per-fiber data for spectropolarimetric work or combined AB extraction for intensity-only work. The calibration fiber allows one to observe a reference Fabry-Perot (FP) simultaneously to science observations and calibrate intra-night drifts; on typical nights, these drifts are  at the $<2$-m/s level. In the present work, we used combined AB extractions and their corresponding C fiber drift files. {In addition to accurate wavelength calibration, it is essential to remove telluric  absorption signatures to minimize their imprint in science data. \texttt{APERO} handles telluric absorption in a two-step process. A first-order correction is done by adjusting a TAPAS \citep{bertaux_tapas_2014} absorption model that only leaves the H$_2$O  and dry  absorption (combining CH$_4$, O$_2$, CO$_2$, N$_2$O and O$_3$) optical depths as free parameters. This leaves percent-level residuals for deep ($>50$\%) absorption features and has the advantage of simplicity. This pre-cleaning is also performed on a set of rapidly rotating hot stars, from which we derive a library of residuals that are correlated against H$_2$O and dry absorption  optical depth to create a  residual model with 3 degrees of freedom per pixel (one term proportional each to the optical depth of H$_2$O and dry components and a constant). This telluric residual model is subtracted from science frames after the TAPAS pre-cleaning. This technique leaves comparable residuals to PCA-based methods \citep{artigau_telluric-line_2014}, but has the advantage of being applied locally; spurious pixels in either the hot star or science observations only affect a handful pixels rather than the entire fit. A representative telluric-cleaned spectrum before and after telluric correction is shown in Figure~\ref{fig:gl699_spectrum}.} Once pRV measurements have been performed, a two-step correction of the RV measurements is necessary. One first accounts for the drifts as measured with the C fiber simultaneously with science observations to account for intra-night drifts. In addition, a zero point correction is applied to the data from the observation of a set of RV-stable calibrators similar to what is done with the SOPHIE spectrograph \citep{courcol_sophie_2015}. The detail of zero point correction will be presented in  a future contribution (Vandal et al., in prep).  The dataset comprises of 871 observations. The median per-point error is 2.57\,m/s with a standard deviation of 2.59\,m/s. We did not bin the individual frames, keeping the 4 exposures within each sequence as individual measurements in the reported time series.

\subsection{HARPS data on Proxima}
We used HARPS \citep{pepe_harps:_2002} publicly available data on Proxima Centauri to assess the performance of the LBL technique in the optical domain. We retrieved the 2D (per-order) spectra and CCF data (M2V mask) from the ESO science archive \citep{delmotte_eso_2006}, and concentrated only on epochs following the fiber upgrade  \citep{lo_curto_harps_2015}, that is after May 29, 2015. A simple telluric correction was performed by adjusting a TAPAS \citep{bertaux_tapas_2014} model to the data. The LBL velocities are then compared in Section~\ref{sec:harps_proxima} to the same data analysed with the \texttt{HARPS-TERRA} and \texttt{NAIRA} template matching algorithms \citep{damasso_low-mass_2020, astudillo-defru_harps_2017}. The full-width at half maximum (FWHM) information, used as an activity indicator and discussed in Section~\ref{sec:proxima_activity}, were drawn from the CCF measurements provided by the ESO science archive.

\section{The Line-by-Line method\label{lbl_method}}
\subsection{Motivation}

One of the main issues with $any$ pRV measurement is the handling of spurious data lines in the spectrum that, if unaccounted for, can bias the measurement well above the typical uncertainties due to the white noise contribution of the data. In the absence of spurious outliers, the RV measurement is conceptually straightforward; one obtains a high-SNR template from the ensemble of observations and performs the summations in Equations \ref{sum_rv} and \ref{err_rv} to derive a velocity and the corresponding uncertainty. In essence, spurious regions in the spectrum are problematic as at some line-of-sight velocities they will randomly align with sharp spectroscopic features such as line wings (i.e., lines of large $\left| \frac{ \partial A_0(i) }{\partial \lambda(i)} \right|$). Regardless of the method, one needs to identify and reject these spurious lines.

Precision radial velocity in the near-infrared is further complicated by a number of issues. First of all, the Earth's atmosphere displays numerous absorption bands that are superimposed on the science data and must be corrected. Irrespective of the telluric subtraction methods used,  one expects an increased noise in the telluric-cleaned regions from the lower effective throughput, as well as systematic effects due to the limitations of models and unaccounted effects in the calibration data. One can reject the domain surrounding deep absorption, but in order to preserve the same domain observed throughout the year, one needs to reject a domain that is up to $\sim64$\,km/s larger than the line itself because of Earth's orbit that adds line-of-sight change of $\pm32$\,km/s for ecliptic targets. The loss is exemplified in \cite{reiners_detecting_2010} that reports a $55$\,\% and $46$\,\% effective loss in wavelength domain due to telluric absorption in $J$ and $H$ bands respectively. As we show in Section~\ref{gains}, this loss can be reduced to nearly zero with the proper data analysis framework.
%

\subsection{The method}
The LBL method is conceptually rather simple; one performs the B01 summation on the spectra in order to derive a velocity in a way that is optimal considering the noise in the data. The subtlety being that this sum is done step-wise; one performs the sum on relatively small domains called `lines'. One then performs a consistency check between the velocity of all lines, and finally average line velocities into a single per-spectrum velocity with the corresponding uncertainty, propagated from the per-line uncertainty. As shown in \citet{dumusque_measuring_2018} and \citet{cretignier_measuring_2020}, the LBL algorithm can be efficient in identifying, and accounting for, lines affected by stellar activity for Sun-like stars in the optical domain. Here, the two-step process allows one to perform a very efficient rejection of lines polluted by a number of artifacts, both known (telluric absorption artifacts, detector defects, sky emission lines) and unknown effects. The LBL method is not immune to other effects such as the propagation of uncertainties in the wavelength solution. This type of uncertainty is different in nature from RV uncertainties induced by artifacts in the spectra and will affect any pRV method equally.

\subsection{Gains from domain splitting}
\label{gains}

When obtaining a single RV measurement for a given spectrum, it is problematic to determine whether the measurement has been significantly biased by unknown residuals, either from uncertainties in telluric absorption correction, sky emission subtraction or unknown effects. As radial velocity effects from planets are expected to be achromatic, one can divide the spectrum into a number of sub-domains, each having its own velocity measurement and verify the statistical consistency between these measurements. As a thought experiment, consider a spectrum with a 1\,m/s theoretical RV accuracy, but having a single uncorrected glitch in the data that leads to a $2$\,m/s error in the global RV measurement. If the spectrum is subdivided in $N$ domains with equal RV content, the  RV accuracy of each domain will be reduced by a factor $\sqrt{N}$, but the velocity error induced by the glitch will increase by a factor $N$ within its segment. If the domain is split in $N=100$ segments, the RV uncertainty for each segment will be 10\,m/s, but the segment affected by the glitch will have its velocity offset by 200\,m/s, or a 20\,$\sigma$ outlier that can readily be identified. Furthermore, the cost in wavelength coverage of rejecting the segment with a spurious velocity will be small, scaling down as $N^{-1}$. There is an additional benefit with the increase in $N$: by setting a certain threshold in RV offset relative to all other segments, one sets an upper limit on the discrepancy of a given segment to all others in the spectrum, then the maximum global RV error injected from a glitch scales down as $1/\sqrt{N}$. In the toy model, a 5-$\sigma$ cutoff for segments would set an upper limit at 50\,m/s on any segment relative to others, which in turn, cannot inject an error larger than 0.5\,m/s in the average radial velocity. For a global RV accuracy of $\sigma_{\rm g}$, $N$ segments and a  rejection at the $n_{\sigma {\rm c}}$ level, the largest error a glitch could inject in the global RV is $n_{\sigma {\rm c}} \sigma_{\rm g}/\sqrt{N}$. The idea of splitting a spectrum in numerous sub-domains to identify outliers has also been used before in the context of iodine cell pRV measurements (see section 6.5 in \citealt{hatzes_doppler_2019}).


\subsection{Mathematical framework}
\label{mathematical_framework}

\begin{figure}[!htbp]
    \centering
    \includegraphics[width=\linewidth]{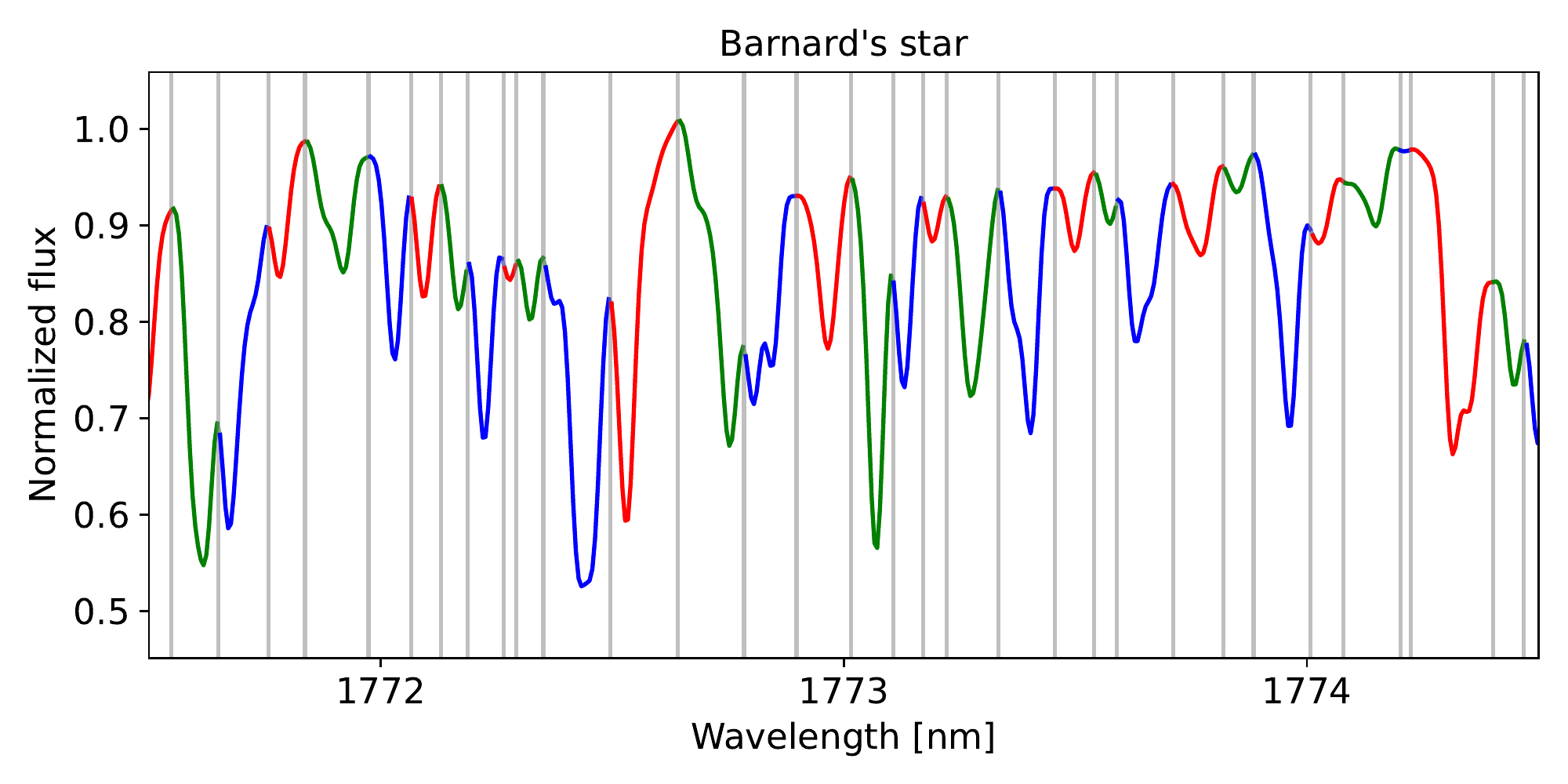}
    \caption{Sample region of the Barnard's star template spectrum with limit of {\it line} domains shown as dashed grey lines.  Color coding is arbitrary and highlights individual lines.}
    \label{fig:mask}
\end{figure}

For the purpose of the LBL algorithm, a line is simply defined as being the domain between two consecutive local maxima in the template spectrum. This has the advantage of simplicity,  allowing the unambiguous empirical definition of lines in dense molecular bands. This method also benefits from being independent of resolution once stellar features are resolved. One can numerically differentiate the template, and find points having a zero derivative and negative second derivative (i.e., local maxima). The noise in the template should be significantly smaller than the point-to-point flux change in order to avoid identifying noise excursions as lines. Figure~\ref{fig:mask} illustrates the positions of lines for a representative region of the SPIRou Barnard's star spectrum template.  To avoid having very narrow lines that may be spurious in a nearly flat region of the spectrum,  we smooth the template with a boxcar kernel of 1 FWHM prior to finding the local maxima.

\begin{figure*}[!htbp]
    \centering
    \includegraphics[width=\linewidth]{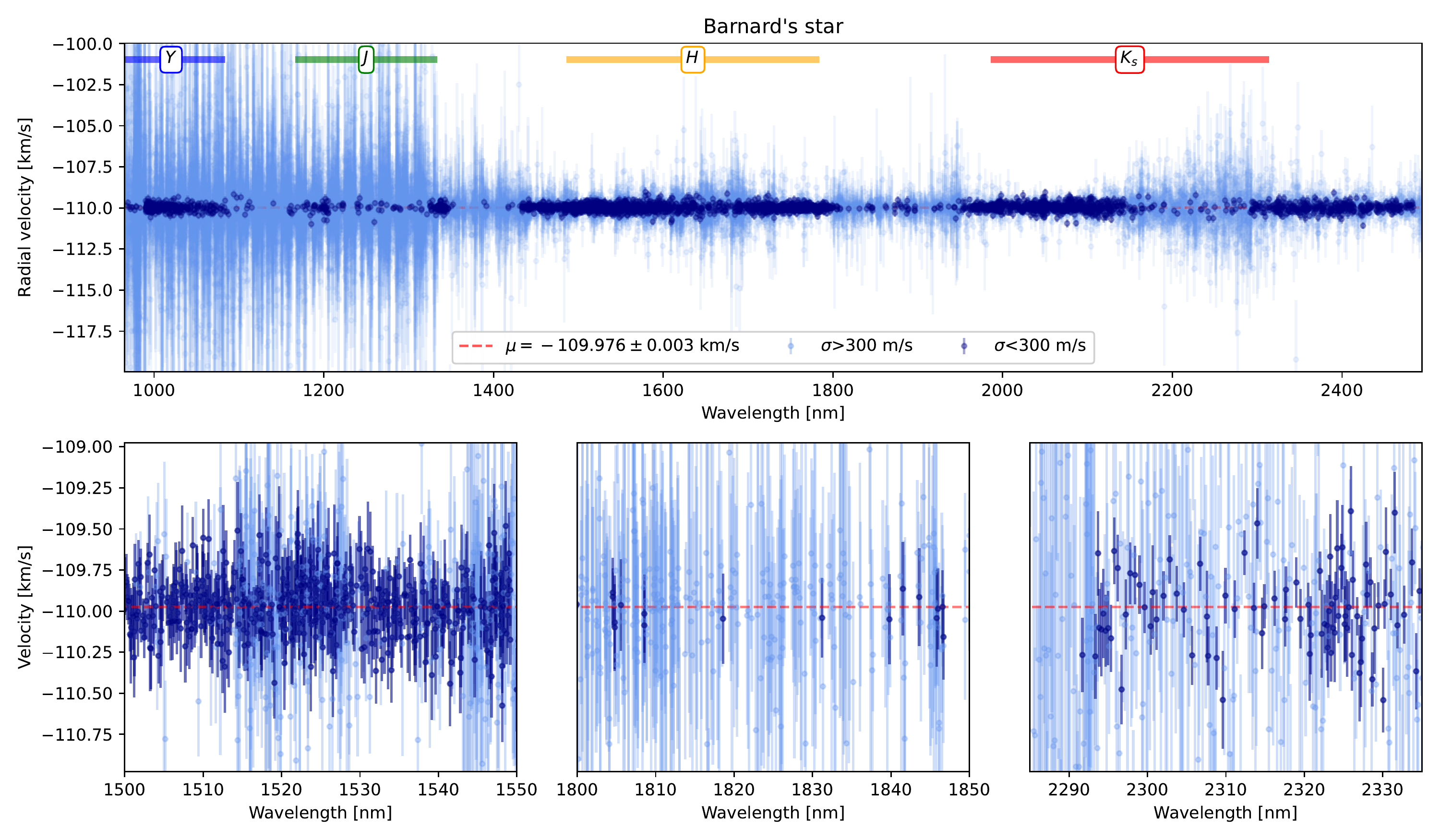}
    \caption{Sample LBL velocity spectrum for Barnard's star showing the scatter of velocity measurements over a wide velocity range ($\pm10$\,km/s of mean velocity; upper panel); lines with  uncertainties smaller than 300\,m/s are shown in dark blue to improve readability. Lines with uncertainties $<$\,300\,m/s are mostly concentrated within the $H$ band and a large part of the $Y$ and $J$ bands are dominated by lines with km/s-level  RV uncertainties. Lower panels show the same plot zoomed on one of the domain with the highest density of high-accuracy lines (1500-1550\,nm; compare with RV accuracy in Figure~\ref{fig:rvcontent}), a domain strongly affected by telluric absorption but with significant RV content (1800-1850\,nm) and the CO bandhead (past 2290\,nm). Parts of the domain, such as the redder half of $K$ band and most of $J$ band, have very few, if any, high-quality lines. These regions correspond to domains where lines are either shallow ($K$) or broader (e.g., $J$; see \citealt{artigau_optical_2018}) than they are in $H$ band. This is also seen in Figure~\ref{fig:rvcontent}. These domains contribute very little to the overall RV budget. The large number of high-quality lines past 2290\,nm (lower-left inset) corresponds to the CO band heads.}
    \label{trumpet}
\end{figure*}

\begin{figure}[!htbp]
    \centering
    \includegraphics[width=\linewidth]{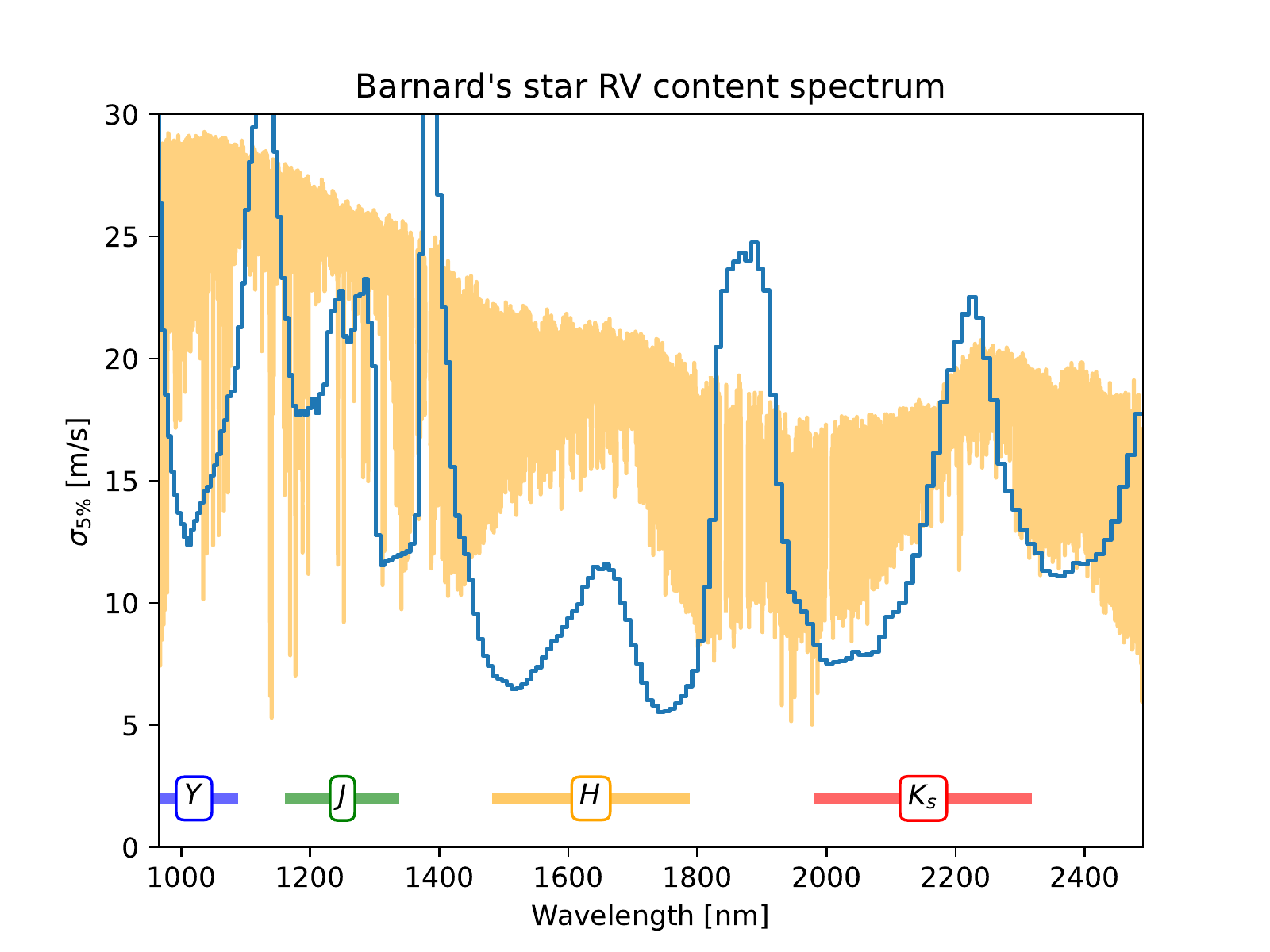}
    \caption{Radial velocity accuracy achieved for a \hbox{$\Delta\lambda/\lambda=5$\%} running window (blue line). As demonstrated in \citep{artigau_optical_2018}, the $H$ band contains the domain with the highest RV content for the near-infrared domain. Remarkably, a single bin centered at 1.75\,$\mu$m would allow a 5\,m/s RV accuracy, only a factor of $\sim$2 worse than the full domain of SPIRou (2.53\,m/s) for the observation considered. The full spectrum of Barnard's star is shown as an orange overplot.
    }
    \label{fig:rvcontent}
\end{figure}

Given a template and a line domain, one computes a per-line velocity using Equation \ref{sum_rv} and its corresponding uncertainty with Equation \ref{err_rv}. One then derives the {\it velocity spectrum} (see Figure~\ref{trumpet}); in the case of Barnard's star, there are $\sim16\,000$ velocities with widely varying uncertainties, the best lines have uncertainty at the $50$\,m/s level while the worst ones have uncertainties at the many km/s level. The global RV precision for the sample spectrum discussed in Figures~\ref{fig:gl699_spectrum}, \ref{fig:mask}, \ref{trumpet}, and \ref{fig:rvcontent} is 2.53\,m/s.
There are a number of interesting features in the velocity spectrum. It is notable that a significant fraction of the RV content is retrieved from the telluric absorptions {\it outside} of the nominal $YJHK$ bandpasses. The lower-middle panel of Figure~\ref{trumpet} shows lines between $H$ and $K$ band, at $\sim1820$\,nm, that have a good RV accuracy (some at $<300$\,m/s/line) without obvious outliers to the mean velocity. Uncertainties are strongly chromatic, being significantly smaller in $H$ and $K$ band than blueward of 1400\,nm. The fact that the best domain in the near-infrared lays long-ward of 1400\,nm was documented by \citet{artigau_optical_2018} and \citet{figueira_radial_2016}. By performing a running weighted mean (see Appendix~\ref{appendix_A}) on the velocity spectrum over a $\Delta\lambda/\lambda = 5$\,\%, one can directly and precisely measure this RV content, as shown in Figure~\ref{fig:rvcontent}.

\subsection{Iterations of the LBL algorithm}
The B01 equations are only valid in the regime for which the higher order derivatives are vanishingly small compared to the first-order derivative in Equation~\ref{eq:b01}. This is analogous to the Newton algorithm to find the roots of a function by representing a function as a Taylor series expansion of the first order. 
This assumption is not always valid for pRV series, and one should use the LBL algorithm iteratively to satisfy this requirement. For a given observation, we spline the template $A_0$ to the expected systemic velocity of the star plus its BERV offset, derive $\delta V$ and $\sigma_V$ values, and if $\delta V$ is not consistent with zero (in practice $|\delta V| < 0.1\,\sigma_V$), we update the input velocity of $A_0$ and recompute  $\delta V$ and $\sigma_V$. The LBL algorithm has a `capture velocity' that is about the FWHM of the effective line profile, which is typically 5\,km/s for mid-Ms in the near-infrared \citep{artigau_optical_2018}. The algorithm fails to converge if the input velocity is off by tens of km/s, e.g., not accounting for the BERV in the initial splining of $A_0$. If the systemic velocity of the target is unknown, a first estimate should be determined beforehand, at the sub-FWHM level, with other numerical methods that can scan over large velocity ranges (e.g., CCF). For the LBL, the absolute velocity measurement of a template is described in Section~\ref{sec:absolute}. For RV-stable targets (with planetary and activity signals at the few m/s level), the algorithm converges in one or two steps. To test the convergence in the presence of a large RV signal, we revisited the \hbox{TOI-1278} dataset presented in \citet{artigau_toi1278_2021} where an 18\,M$_{\rm Jup}$ brown dwarf imparts a $K\sim2300$\,m/s RV signal on it M0V host. The RV excursions for the system are at about half of the FWHM of stellar lines. For TOI-1278, the algorithm typically converges in 3 steps, with a typical residual uncertainty that decreases by a factor  $\sim30$ between steps. The per-epoch RV uncertainty with the LBL is 9.4\,m/s, which is a two-fold improvement compared to the values published values which were measured with a CCF (see Table~1 in \citealt{artigau_toi1278_2021}). The TOI-1278 Keplerian fit values are consistent with those published albeit with significantly reduced uncertainties; this target is currently the subject of ongoing monitoring with SPIRou and its extended RV monitoring will be presented in a future publication.

\subsection{Outliers}
One key point in the LBL analysis is the filtering of problematic lines, and therefore the identification and abundance of outliers in the velocity spectra. Assuming that uncertainties are Gaussian in the input spectrum and properly accounted for, the B01 framework should lead to a normal distribution of uncertainties for all lines. In practice, we show that this is nearly the case out to 2\,$\sigma$, with a small fraction of $>$\,5\,$\sigma$ outliers (see Figure~\ref{fig:sigma_dist}). It is notable that outliers in the velocity spectrum do not specifically appear in telluric bands as one could expect (see Figures~\ref{trumpet}, \ref{fig:sigma_dist} and Appendix~\ref{appendix_A}).  This is suggestive that the \texttt{APERO} telluric subtraction method is effective. The presence of a small number of outliers in the line velocities implies that one must use a form of sigma-clipping. We adopt a simple mixture model \citep{hogg_data_2010, bishop_pattern_2006} to estimate the mean velocity, in which there are two distributions, one for valid lines that follow a normal distribution corresponding to the uncertainties as derived from the B01 framework and having a common velocity, and the other, broader, distribution corresponding to outliers. Appendix~\ref{appendix_A} provides the details on the numerical implementation. The resulting per-epoch velocity is $\overline{v}$ with the corresponding uncertainty, $\overline{\sigma}$.

The LBL algorithm comes with a notable advantage relative to a number of other pRV methods regarding the RV content: it does not require the rejection of any wavelength domain to account for yearly BERV variations. In order to preserve the wavelength domain through the year, we accept that the domain will change with the yearly line-of-sight velocity variations and reject lines if they are found to be discrepant. This significantly increases the effective RV content of a given observation and basically sets us in the $condition\,3$ of \citealt{figueira_radial_2016} (only increase in  variance due to the lower transmission is accounted for) rather than the very conservative $condition\,2$ therein (rejection of all lines within $\pm30$\,km/s of any absorption) and gets close to the RV accuracy that would be possible without any telluric absorption over the same domain ($condition\,1$).  Furthermore, unlike several template matching methods, the LBL algorithm does not require any free parameters; there is no threshold in the equivalent of sigma-clipping, this being handled in a statistically consistent way (see Appendix~\ref{appendix_A}).

\subsection{Debiasing}
One potential issue with the LBL method is that the actual list of valid lines will vary slightly from spectrum to spectrum. If a line has a mean velocity through the entire time series that is slightly offset from the velocity spectrum mean, then having it rejected at some epochs will lead to a bias. One needs to account for the varying line list. This step is done by constructing a 2D matrix, with N lines along one axis and M velocity spectra along the other axis; this matrix has gaps that vary between spectra.   One needs to construct two 1D vectors, one along the N and one along the M axis that correspond to the per-line bias and per-epoch velocity. Debiasing is done iteratively; one first computes a per-epoch mean velocity using the mixture model in  Appendix~\ref{appendix_A} and then computes a per-line mean velocity after subtracting the per-epoch mean velocity. This allows to determine the mean per-line velocity offset relative to the mean velocity of all lines. This offset is then subtracted from each line over the entire time series prior to computing the final $\overline{v}$ and $\overline{\sigma}$. This process could be iterated further, but in practice a single iteration is sufficient for velocity to converge well within $\overline{\sigma}$. The debiasing implies that the addition of new data will very slightly affect the RV measured in previous observations as the debiasing model will improve with time. In practical cases, this is a second-order effect that is smaller than the per-epoch RV uncertainty. The debiasing provides a significant improvement in the time series point-to-point scatter; without it, the RMS of the Barnard's star sequence  (2.59 m/s), presented in Section~\ref{demonstrations}, increases by 50\,cm/s.

\subsection{Template construction and usage}

The template spectrum is constructed from the median combination of BERV-registered spectra after telluric correction. The template construction is done with co-added, stitched, spectral orders rather than  per-order spectra. This allows for a proper accounting of wavelength overlap between consecutive orders. Figure~\ref{fig:gl699_spectrum} shows a small domain of Barnard's star SPIRou spectrum, with the telluric correction applied to generate the telluric-corrected and template spectrum. Template creation is part of the \texttt{APERO} outputs and a stand-alone template creation is provided with the LBL tools (See Appendix~\ref{code_accessibility}).

Considering that numerous lines and spectral features of interest have typical depths at the $>$5\% level relative to the surrounding mean flux (a true continuum cannot be defined in the rich molecular bands of M dwarfs; see Figure~\ref{fig:mask}), one needs to construct a template that has a sufficiently good SNR for identified lines to be true spectroscopic features rather than noise excursions. In practice, a template with SNR $>200$ is sufficient to unambiguously identify spectral features and define `lines'. Considering that pRV time series typically have tens of observations, each with SNR $>50$, this condition is generally met as the combined SNR of the templates scales as the square root of the number of observations. One further requirement in constructing the template is that any large RV shift from companions should be accounted for if the $\frac{\partial A_0(i)}{\partial \lambda(i)}\frac{ \lambda(i) \delta V}{c}\ll \sigma(i)$ condition is not met. This condition is generally met for planetary companions with signals below the $\sim$100\,m/s level.  If no template is available or a strong planetary signal is present, one can use a template from a star of a similar spectral type and rotational velocity. We computed the LBL sequence on Barnard's star detailed in Section~\ref{spirou_barnard} with the template constructed from GJ\,1012 observations; both stars are M4V and slow rotators and the latter had a large number (449) of observations obtained through the SPIRou Legacy Survey \citep{donati_spirou_2020}. The resulting time series is similar to the one obtained with Barnard's star's template, albeit with a $\sim$20\% worse point-to-point scatter. While we have not systematically explored the degradation of pRV performances for all sub-type combinations, time series obtained with a model matching within $\pm$2 stellar sub-types generally have performances similar to those obtain with an object's own template.

\subsection{Absolute velocity measurement}
\label{sec:absolute}
 The LBL algorithm is by nature differential as it measures a velocity difference between a template and a given science spectrum. It is often desirable to have an absolute velocity for a given observation, so we measure the systemic velocity of our templates and add it to the velocity difference between an observation and the corresponding template. To measure the velocity of templates, we find the lines using the method described above and obtain a CCF by cross-correlating this line list with a PHOENIX synthetic spectrum matching the star's temperature; these spectra are drawn from the Goettingen spectral library \citep{husser_new_2013}. The peak of the CCF using an empirical line position onto a synthetic spectrum is expected to be opposite of the star's systemic velocity. By using different templates onto the same time series, we can estimate the typical errors on the systemic velocity. We used the Barnard's star sequence described here and measured its velocity with templates from 11 stars within 100\,K (GJ1012, GJ1103, GJ1105, GJ1151, GJ1289, GL15B, GL169.1A, GL445, GL447, Barnard's star, GL725B). The systemic velocity of Barnard with it's own template is $-109\,974$\,m/s, while the dispersion of systemic velocities derived using all templates is 37\,m/s. We therefore estimate that the typical uncertainties on systemic velocities from the LBL are at that level for high-SNR observations. We note that some systemic effects such as uncertainties in the line positions of models or limits to our knowledge of the gravitational redshift are common to all stars and not accounted for in this error estimate.

\subsection{High-pass filtering  data}
\label{high_pass_filter}
One problem in subtracting a template from a science observation is the proper scaling of the template in amplitude. One can leave each line with an arbitrary scaling factor, as is done in \cite{dumusque_measuring_2018}, or adopt a coherent amplitude over a larger spectral domain. Furthermore,  infrared astronomical arrays display persistence; this is an effect where a ghost image of the previous observation is detected at reduced amplitude in subsequent images, even after detector reset (e.g., \citealt{artigau_h4rg_2018, baril_characterization_2008}). In nIR high-resolution spectroscopy, persistence mostly displays as a dilution of the line depths as it adds a continuum contribution that is distinct from that of the target. Modelling persistence is challenging as it requires knowledge of the entire time series of the night, including calibrations. In order to avoid problems linked to the dilution of the continuum in nIR observations both the science spectrum and the template are high-passed by subtracting a low-pass filter over a 500\,km/s running window. The high-pass filtering is performed by subtracting the running mean of a window of a given width centered on the wavelength of interest. We explored windows ranging down to 50\,km/s and this parameter has minimal impact on the end-result.  This effectively removes the persistence contribution to the continuum level, an issue that has shown to be problematic in the case of observing fainter M dwarfs with SPIRou, in particular TRAPPIST-1, immediately after either telluric standards or transit spectroscopy targets that compound a bright target with a long staring, worsening persistence effects. After high-passing, the template is multiplied by a scaling factor to match the amplitude of the science spectrum through a least-square adjustment. This is done per-order, allowing for slight large-scale chromatic losses  between science observations. Chromatic losses could arise from varying scattering properties of the atmosphere, coating degradation in telescope optics or varying fiber coupling efficiency under varying seeing, the seeing profile being mildly chromatic \citep{fried_optical_1966}. In typical SPIRou observations, the median $J$-to-$K$ flux ratio typically varies by up to 15\% on monthly timescales.

\subsection{Activity indicator}
\label{sec:activity_indicator}

The B01 framework corresponds to a projection of the residuals between a given spectrum and the corresponding template onto the first derivative of the template with wavelength. One can extend this framework, by letting the spectrum $A(i)$ line profile width vary by infinitesimal amount $\delta w$, relative to the mean width $w$ of the template $A_0(i)$:
\begin{equation}
A(i) - A_0(i) = \frac{\partial A_0(i)}{\partial \lambda(i)}\delta\lambda(i) + \frac{\partial A_0(i)}{\partial w(i)} \delta w(i)
\label{eq:b01extended}
\end{equation}

In this context, a spectrum as being the sum of well separated Gaussians having the same widths leads to an interesting analytical solution. Admittedly M dwarfs spectra are much more complex with numerous overlapping features, but this framework does provide insight into the physical interpretation of the projection of residual onto higher-order derivatives. For a Gaussian line having a width $w_{\rm g}$ centered at $\lambda_0$, the equivalence between the first derivative against $w_{\rm g}$ and the second derivative with respect to wavelength is given by:
\begin{equation}
\frac{\partial}{\partial w_{\rm g}}\left(\frac{e^{-0.5\left(\frac{\lambda-\lambda_0}{w_{\rm g}}\right)^2}}{w_{\rm g}}\right) = w_{\rm g} \, \frac{\partial^2}{\partial \lambda^2}\left(\frac{e^{-0.5\left(\frac{\lambda-\lambda_0}{w_{\rm g}}\right)^2}}{w_{\rm g}}\right)
\label{eq:equivalence}
\end{equation}

In other words, changes in the line profile width ($\delta w$) from spectrum to spectrum may be approximated as a second derivative term in wavelength, as long as $\delta w / w \ll 1$.  In the case of rich molecular bands, this is an approximation as a line width cannot be rigorously defined, but it would be strictly true for isolated Gaussian lines with a $w_{\rm g}$ width. By taking Equations \ref{eq:b01extended} and \ref{eq:equivalence}, 
we have:
\begin{equation}
    A(i) - A_0(i) = \frac{\partial A_0(i)}{\partial \lambda(i)}\delta\lambda(i) + \frac{\partial^2 A_0(i)}{\partial \lambda(i)^2} w_{\rm g}(i)\delta w_{\rm g}(i)
    \label{eq:bouchy_2d}
\end{equation}

This correspondence between the second derivative of the spectrum and the width for Gaussian lines was previously discussed in details in \citealt{zechmeister_spectrum_2018} (Section 3.2 and Appendix\,B therein); they propose an activity indicator, the differential line width ($\textrm{dLW} \equiv\sigma\Delta\sigma$), where $\sigma$ corresponds to the equivalent width of Gaussian lines expressed in velocity unit and $\Delta \sigma$, the variation of this width. Taking Equation~\ref{eq:bouchy_2d} for processes with $\frac{\partial A_0(i)}{\partial \lambda(i)} = 0$ (e.g., symmetric line profile changes such as a broadening), we have: 
\begin{equation}
\frac{\textrm{dLW}(i)}{c^2} = \frac{A(i) - A_0(i)}{\lambda(i)^2\frac{\partial^2 A_0(i)}{\partial\lambda(i)^2} }
\end{equation}

Following the same formalism as for Equations~\ref{eq:noise} and \ref{sum_rv}, one can derive a line-by-line dLW, which can then be averaged out in an outlier-resilient way (as described in Appendix~\ref{appendix_A}). The dLW indicator has m$^2$\,s$^{-2}$ units and scales with the variation in the line profile FWHM, a metric widely used in the literature to characterize the activity level of a star. Considering that a Gaussian line has a $\textrm{FWHM} = 2\sqrt{2 \ln 2}\,w_{\rm g} \approx 2.35 w_{\rm g}$, one obtains:
\begin{equation}
\textrm{dLW} = \frac{\textrm{FWHM} \times \Delta \textrm{FWHM}}{8 \ln 2}
\end{equation}

However, one caveat in converting a dLW measurement into $\Delta$FWHM is that the knowledge of the effective FWHM of the template is required. One can derive this value by using a CCF to allow a direct comparison with FWHM changes detected through other methods. If dLW is used as an activity indicator alone, one can use it directly for period searches and detrending. This is demonstrated on the Proxima Centauri data in Section~\ref{sec:proxima_activity}. The weighted mean in velocity and dLW are computed independently for the sake of simplicity; this implies that a given point may be an outlier within one distribution but not in the other. We tested a joint rejection, and the impact is minimal, typically well below 0.1-$\sigma$ for either value.


One notable characteristic of the dLW is that a variation in the instrument line-spread-function profile has the same impact regardless of the stellar FWHM. This is not the case for changes in the FWHM as measured with a CCF; the larger the FWHM, the smaller the impact of a change in the instrumental profile. This implies that long-term monitoring of activity can be corrected for instrumental profile changes directly by using the dLW, as measured on a set of calibrator stars or with calibration data such as the Fabry-Perot spectrum used to determine the wavelength solution.

\section{Demonstrations of the LBL method\label{demonstrations}}
The LBL algorithm allows the measurement of Keplerian signals at the few m/s with m/s-level uncertainties in SPIRou datasets, as presented in recent work. It has been used to measure \hbox{TOI-1759\,b} mass through a measurement of a pRV signal of \hbox{$K=2.3\pm0.7$\,m/s} \citep{martioli_toi-1759_2022-1}. Similarly, it has been used to measure the mass of the sub-Neptunes orbiting TOI-2136 and TOI-1452 (\citealt{gan_tess_2022}; Cadieux et al., submitted), respectively corresponding to signals at the $K=4.2\pm1.4$\,m/s and $3.5\pm0.9$\,m/s level. These discoveries are based on individual measurements at the 6 to 10\,m/s with typically 4 measurements (one polarimetric sequence) per epoch, which are limited by the modest SNR obtained on these relatively faint targets ($H=9.6-10.0$) and do not represent a fundamental noise floor for the technique. Here, we present a detailed analysis of the multi-year time series obtained on Barnard's star at IR wavelength and Proxima in the optical (respectively with SPIRou and HARPS), focusing on the statistical properties of these datasets in comparison with other pRV measurement methods.

\subsection{LBL vs.\ CCF on Barnard's star} \label{spirou_barnard}
Barnard's star has been observed regularly since the first light of SPIRou, and it serves as a benchmark object for the development of the SPIRou data analysis tools. The star is known to display little activity due to its slow rotation and thick-disk age. It is also among the brightest M dwarfs in the sky due to its proximity (1.83\,pc, \citealt{gaia_collaboration_gaia_2021}).
Furthermore, the recent claim of a Super-Earth  on a \hbox{233-day} orbit \citep{ribas_candidate_2018} and the subsequent attribution of the pRV signal to activity \citep{lubin_stellar_2021} make it a compelling target. We defer the full analysis of the Barnard's star RV time series to a future publication (Vandal et al., in prep.), and present a summary.  

\begin{figure}[!htbp]
    \centering
    \includegraphics[width=\linewidth]{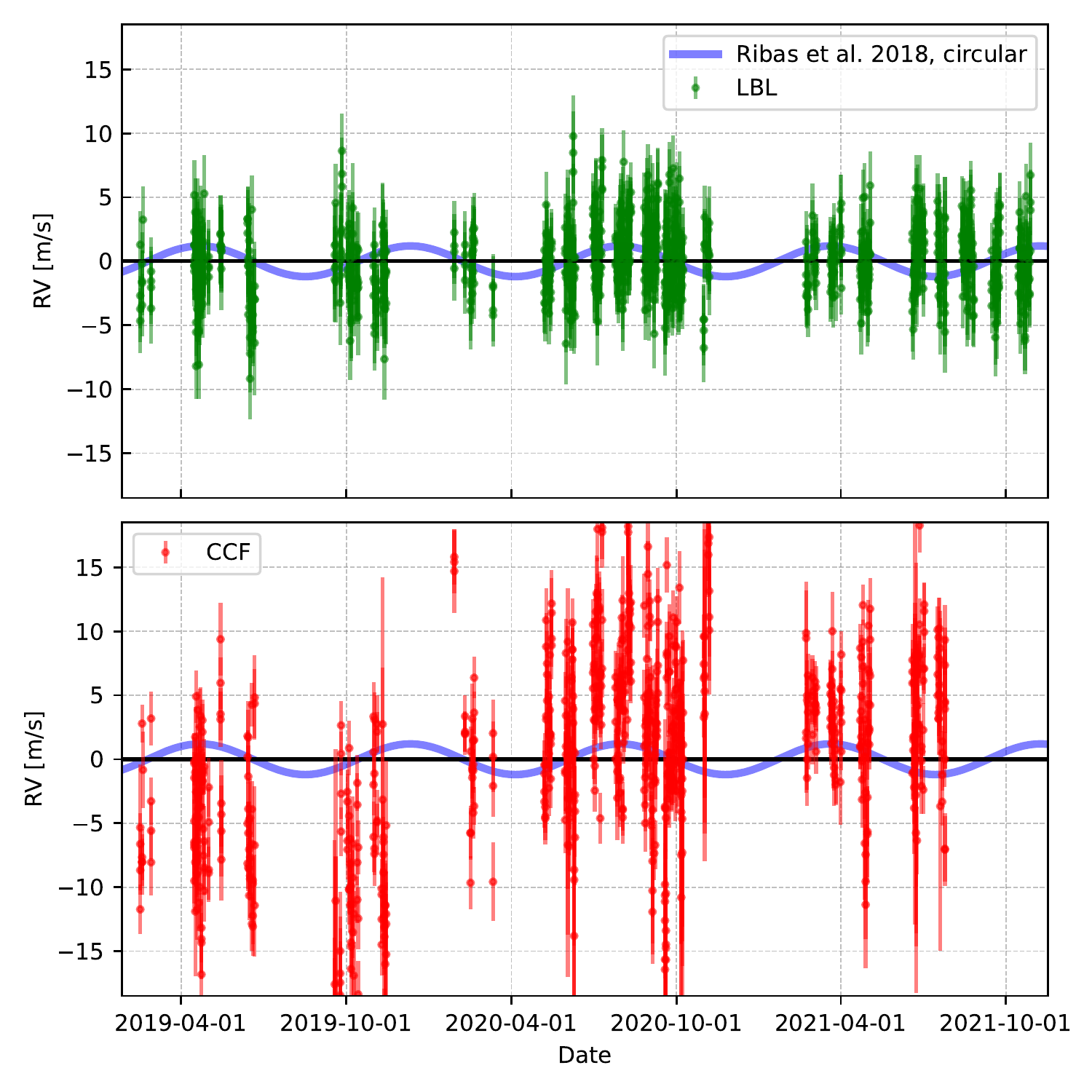}
    \caption{Radial velocity time series of Barnard's star with the LBL method and the standard SPIRou CCF.  No activity correction has been performed. The median LBL RV uncertainty is 2.57\,m/s, only slightly smaller than the point-to-point dispersion 2.59\,m/s. The \cite{ribas_candidate_2018} circular orbital solution is shown. The data strongly favors the scenario for which the radial velocity is constant compared to that with the candidate planet with the reported solution (see Section~\ref{spirou_barnard}).}
    \label{fig:lbl_vs_ccf}
\end{figure}

\begin{figure}[!htbp]
    \centering
    \includegraphics[width=\linewidth]{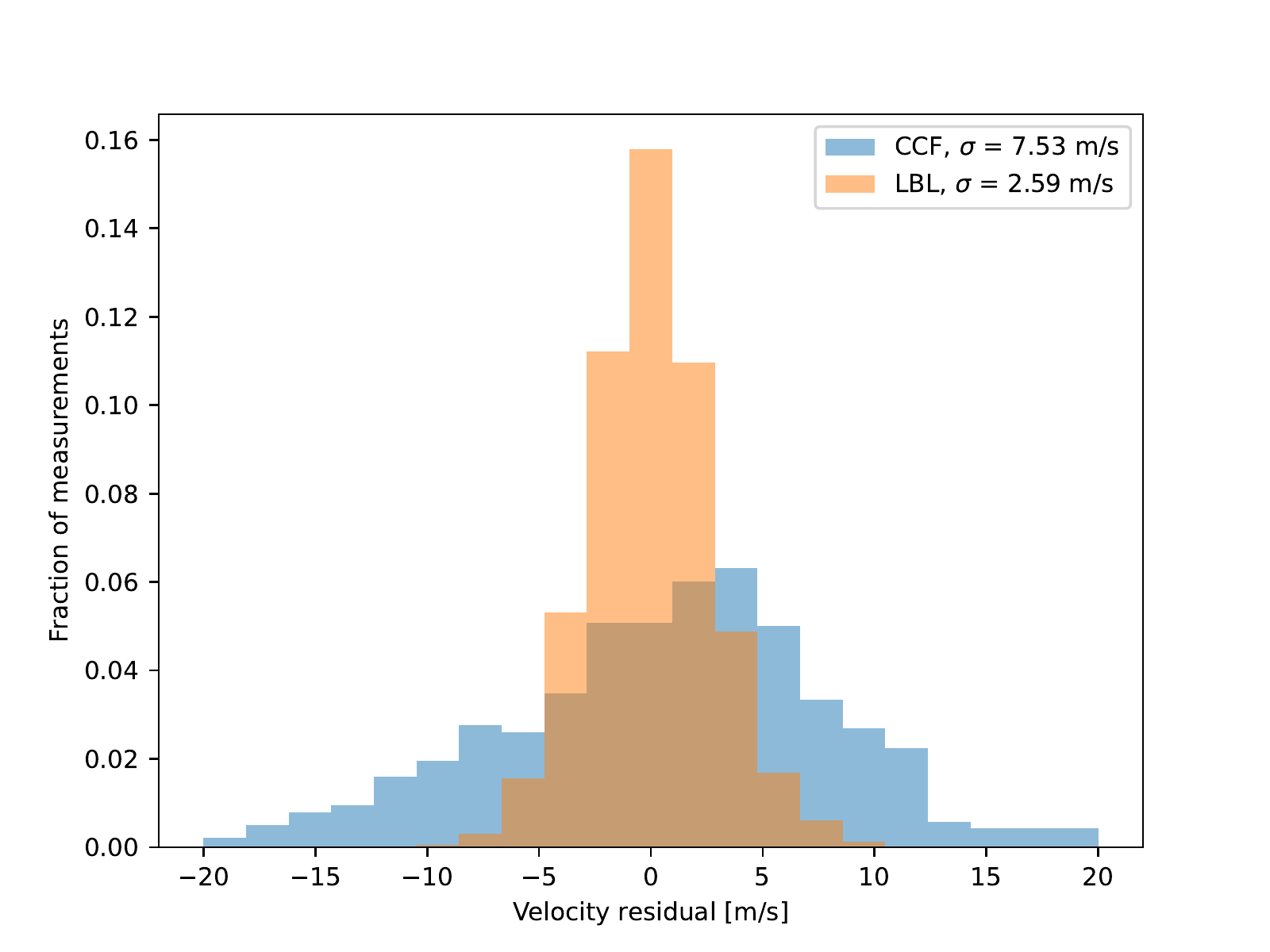}
    \caption{Histogram of radial velocities relative to the median velocity for Barnard's star observed with SPIRou. LBL values have a standard deviation of 2.59\,m/s for a 2.57\,m/s median uncertainty. CCF values have a 7.53\,m/s standard deviation.}
    \label{fig:histo_ccf_lbl}
\end{figure}

We computed Barnard's star's complete RV time series with the LBL and the CCF methods. The CCF RVs were extracted using the \texttt{spirou-ccf}\footnote{\href{https://github.com/edermartioli/spirou-ccf}{github.com/edermartioli/spirou-ccf}} package (see \citealt{martioli_toi-1759_2022-1} for details). The resulting RV curves are presented in Figure~\ref{fig:lbl_vs_ccf}, with the circular Keplerian solution from \cite{ribas_candidate_2018} over-plotted. As shown in the velocity histogram in Figure~\ref{fig:histo_ccf_lbl}, the LBL algorithm provides twice less point-to-point scatter (RMS~= 2.59\,m/s), compared to the CCF derived velocities (RMS~= 7.53\,m/s). The LBL algorithm returns RV uncertainty estimates (2.57\,m/s) that are consistent with the noise level as determined by subtracting a template from the data, but this noise level is, itself, significantly higher than what would be expected from Poisson statistics in the $H$ and $K$ bands. The underlying reason for this excess noise is still being investigated, but modal noise (\citealt{blind_few-mode_2017}, Gaisn\'e et al., in prep.) is one of the main plausible causes, which would explain the worsening at longer wavelengths (i.e., a fiber has fewer modes at longer wavelengths and a fractionally larger modal noise). With only Poisson noise, one would expect an RV accuracy at the 1.0\,m/s level.

Given the 2.59\,m/s dispersion in our Barnard's star LBL time series spanning over 2.7\,years and typical uncertainty of 2.57\,m/s, one would expect the candidate planet unveiled by \citealt{ribas_candidate_2018} ($K = 1.20 \pm 0.12$\,m/s) to dominate the excess noise budget. We conducted a statistical analysis to objectively determine whether this signal is detected in our LBL data. First, we computed the Bayesian log-evidence ($\ln Z$) using the \texttt{nestle} \citep{barbary_nestle_2021} nested sampling algorithm for a flat model, i.e., systemic velocity and extra white noise terms only, and for either of \cite{ribas_candidate_2018} circular and eccentric solutions. Both planetary models are strongly rejected by our data with a difference in log-evidence of $\Delta \ln Z = -23.9$ for the eccentric and $\Delta \ln Z = -33.0$ for the circular solution, compared to the flat solution. This result implies that other sources of noise (e.g., instrumental, stellar jitter, planets) are needed to explain the observed dispersion, those will be addressed in Vandal et al., in prep., with no evidence of the claimed super-Earth in our dataset. Then, we explored the possibility of a 233-day planet, but with a different semi-amplitude $K$ than the one reported in \cite{ribas_candidate_2018}. For this, we generated Keplerian models with \texttt{radvel} \citep{fulton_radvel_2018} and sampled the posterior distribution of $K$ with a Markov chain Monte Carlo (MCMC) Bayesian framework using \texttt{emcee} \citep{foreman-mackey_emcee_2013}. We employed a uniform prior distribution $\mathcal{U}(0, 5)$ for $K$, while the orbital parameters (period and time of periastron) were following a Gaussian prior with standard deviation corresponding to their uncertainty reported in \cite{ribas_candidate_2018}. The plausible values for $K$ (posterior distribution) were incompatible with $K = 1.20 \pm 0.12$\,m/s at the $>$5\,$\sigma$ significance level. Our analysis therefore strongly disfavors the existence of a 233-day planet around Barnard's star with a minimal mass similar to the candidate super-Earth revealed by \cite{ribas_candidate_2018}.

\subsection{LBL vs.\ template matching on Proxima Centauri}\label{sec:harps_proxima}

While the LBL algorithm has been developed with infrared measurements in mind, we tested it on publicly available HARPS data on Proxima Centauri. In addition to being our nearest stellar neighbour, the star is known to host an Earth-mass companion close to its habitable zone \citep{anglada-escude_terrestrial_2016}. More recently, a second close-in terrestrial planet was identified by \citet{faria_candidate_2022}. This well-studied star provides an interesting benchmark for pRV measurement techniques. 

The LBL radial velocities of Proxima were computed using a template generated from all HARPS spectra collected after the fiber change. We compare below the LBL analysis of Proxima with the \citet{damasso_low-mass_2020} publicly available radial velocities derived with the \texttt{HARPS-TERRA} template matching algorithm. We also computed the velocity of the time series with \texttt{NAIRA} (see \citealt{astudillo-defru_harps_2017}, section 3.1), an algorithm to compute RVs through the template matching approach.

 The \texttt{NAIRA} algorithm Doppler shifts all spectra to the stellar frame to compute a median spectrum (stellar template), which is then used to obtain a telluric template from the median of spectra divided by the stellar template shifted to the Earth frame. We iterate the procedure to properly disentangle the stellar and telluric components, which in turn depends on the number of spectra acquired in different BERVs. The telluric spectrum is used to mask only the zones affected by tellurics, instead of discarding $\pm30$\,km/s around a given telluric line. The stellar template is shifted in a range of velocities around the star systemic velocity to compute the likelihood for each RV step. The likelihood is maximized to obtain the optimal RV of the star.

 The main difference between \texttt{HARPS-TERRA} and \texttt{NAIRA} comes from the weights applied to each pixel and individual radial velocities from each aperture. \texttt{NAIRA} performs a simple chi-square minimization assuming Poisson statistics, \texttt{HARPS-TERRA} uses weights that are proportional to the residuals; \texttt{NAIRA} computes the minimization on all apertures (orders) at the same time while \texttt{HARPS-TERRA} obtains an order-by-order minimization to apply weights from their dispersion. Furthermore, \texttt{HARPS-TERRA} masks tellurics deeper than 1\% in a synthetic spectrum and \texttt{NAIRA} masks tellurics for which the Doppler information is significantly larger than the Doppler content of a stellar line located at the same wavelength.


This analysis, on a common dataset allows a direct comparison of the LBL method with state-of-the-art template matching algorithms, and an understanding of eventual differences. The differences seen in the velocity put the detection of the candidate planet at long period identified by \citet{damasso_low-mass_2020} in new light. 

The Proxima Centauri data analysed consists of two densely sampled sequences, one in late-January to early-April 2006 and another one in July to September 2017. A third sequence, obtained in March to July 2019, was analysed with the LBL and \texttt{NAIRA} algorithms but has not been analysed by \citet{damasso_low-mass_2020}.

\begin{figure*}[!htbp]
    \centering
    \includegraphics[width=\linewidth]{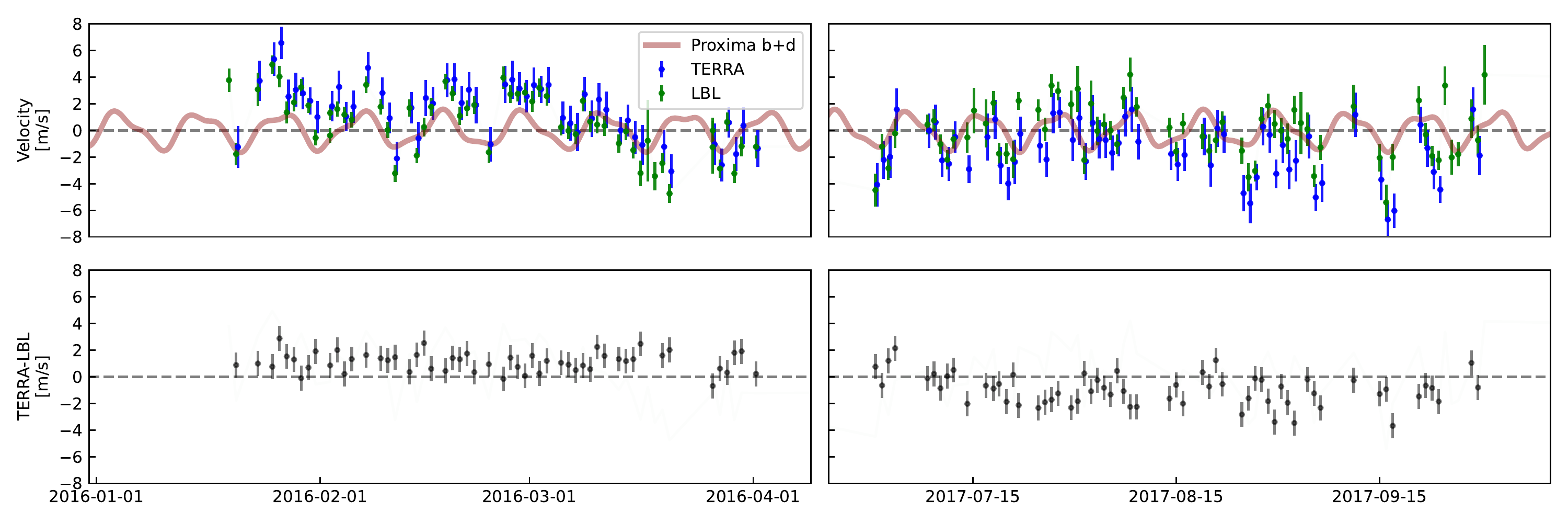}
    \caption{Radial velocity measurements of the HARPS post-fiber change on Proxima Centauri with the LBL and \texttt{HARPS-TERRA} algorithms. The top panel shows both datasets with Proxima\,b and d Keplerian contribution (solution from \citealt{faria_candidate_2022}). The bottom panel shows the difference between LBL and \texttt{HARPS-TERRA}; within each sequence the agreement is excellent (RMS of difference 0.7 and 1.2\,m/s in 2016 and 2017 respectively), but a systematic difference between sequences at the $\sim$3\,m/s is seen. Uncertainties in the difference between \texttt{LBL} and \texttt{HARPS-TERRA} are derived from the per-sequence RMS. One cannot simply add the errors from each method in quadrature as the noise sources in the two are highly correlated; the two used the same input datasets with the same noise, albeit with  different weighting schemes. }
    \label{fig:fit1}
\end{figure*}

\begin{figure*}[!htbp]
    \centering
    \includegraphics[width=\linewidth]{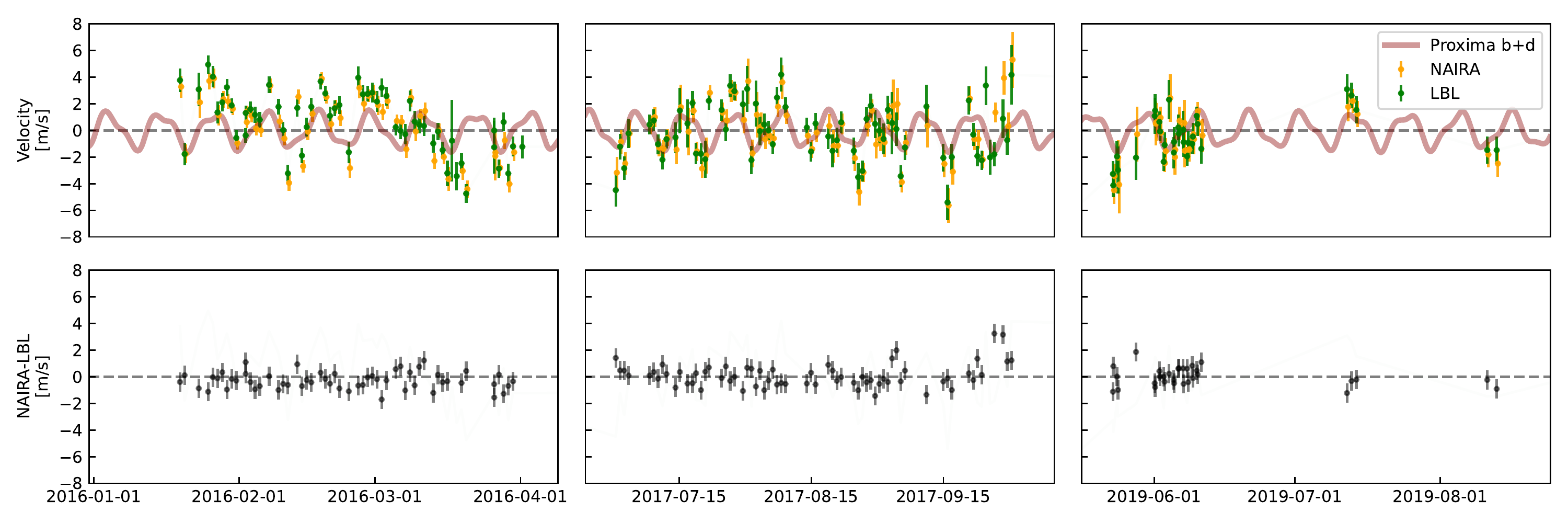}
    \caption{Same as Figure~\ref{fig:fit1}, but for the \texttt{NAIRA} time series. The measurements are in remarkable agreement, with a 0.7\,m/s RMS difference between values. The $\sim$3\,m/s bulk offset between the 2016 and 2017 datasets, as seen in the LBL to \texttt{HARPS-TERRA} comparison, is not present. Overall, the scatter after subtraction of the Proxima\,b and d ephemeris is the same at 1.9\,m/s level for both time series.  }
    \label{fig:fit2}
\end{figure*}

Figure~\ref{fig:fit1} presents the main differences between LBL and \texttt{HARPS-TERRA} radial velocities during the common epochs of 2016 and 2017. If the two extraction methods produce almost perfectly matching RV in timescale of months, the \texttt{HARPS-TERRA} deviate by $\sim$3\,m/s between 2016 and 2017. This trend is absent from the LBL time series. Similarly, Figure~\ref{fig:fit2} provides a comparison between the LBL and \texttt{NAIRA} methods, showing an agreement between the two measurements at the $\sim$70\,cm/s level without any significant difference between sequences.

The most notable difference between the two campaigns is the BERV change; the 2016 sequence corresponds to the largest yearly line-of-sight value (from 16 to 21\,km/s), while the 2017 sequence corresponds with the lowest BERV excursion ($-14$ to $-21$\,km/s over these dates). Any RV extraction method free of systematic effects should yield residuals that are uncorrelated with the BERV once the Keplerian  and activity signals are removed. This can be easily tested on the Proxima data using four RV extraction methods: LBL, CCF, \texttt{HARPS-TERRA}, and \texttt{NAIRA}. To quantify this systematic effect, we define the quantity $K_{\rm BERV}$ as the semi-amplitude of the yearly peak-to-peak signal induced by the correlation between RV residuals and the BERV for that specific target and time sequence. We first removed Proxima\,b and d Keplerian signals in all four datasets using the \cite{faria_candidate_2022} orbital solution. Modelling the activity signal for the four extraction methods was out of scope of this simple analysis. Instead, we estimated how much the activity can contribute in $K_{\rm BERV}$ due to random and finite sampling of Proxima rotation and BERV. We measured the slope of 1000 random realizations of the activity model of the LBL RV time series presented in Section~\ref{section:lnz} with BERV. From this, we derived $K_{\rm BERV}$ values strictly due to activity with a mean of 0.04\,m/s and standard deviation of 0.46\,m/s. We thus added this uncertainty term in quadrature to all four $K_{\rm BERV}$ presented below. As shown in Figure~\ref{fig:berv_vs}, both the CCF and \texttt{HARPS-TERRA} RV residuals show a trend with BERV, i.e., $K_{\rm BERV} = 0.9\pm0.5$ and $1.8\pm0.5$\,m/s, respectively, but this slope is consistent with zero for LBL and \texttt{NAIRA} ($K_{\rm BERV}= 0.3\pm0.5$ and 0.4$\pm$0.5\,m/s). Figure~\ref{fig:berv_vs} strongly suggests that the LBL and the \texttt{NAIRA} template matching method are largely immune to the systematic effects affecting \texttt{HARPS-TERRA} and, to a lesser extent, the CCF measurements. Determining the underlying causes leading to the correlation of velocities with BERV in CCF and \texttt{HARPS-TERRA} data is beyond the scope of this paper, but we note that this behavior is expected for sub-optimal telluric line corrections. While developed in the context of nIR observations, this analysis illustrates the benefit of the outlier statistical rejection that the LBL allows at optical wavelengths. The differences between these models cannot be attributed to differences in zero point or wavelength calibration as they all use the same wavelength grid provided by the HARPS pipeline.

\begin{figure}[!htbp]
    \centering
    \includegraphics[width=1\linewidth]{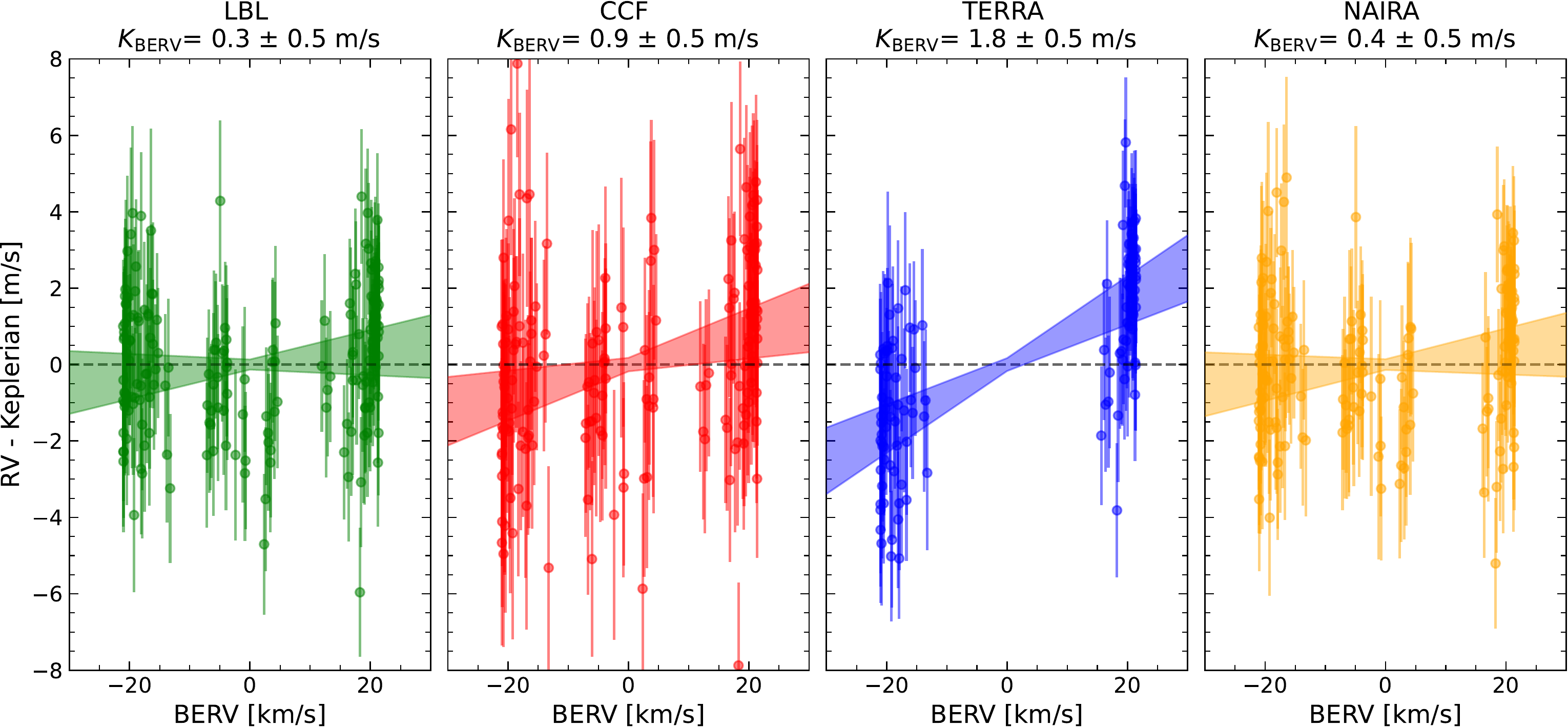}
    \caption{Radial velocity residuals as a function of BERV after subtracting Proxima\,b and d signals using the \citet{faria_candidate_2022} orbital fit. The \texttt{HARPS-TERRA} values display a $K_{\rm BERV}$ signal at the 1.8\,m/s level while it is consistent with zero for the LBL and \texttt{NAIRA} analysis. The CCF data shows an intermediate $K_{\rm BERV}$.}
    \label{fig:berv_vs}
\end{figure}

\begin{figure}[!htbp]
    \centering
    \includegraphics[width=\linewidth]{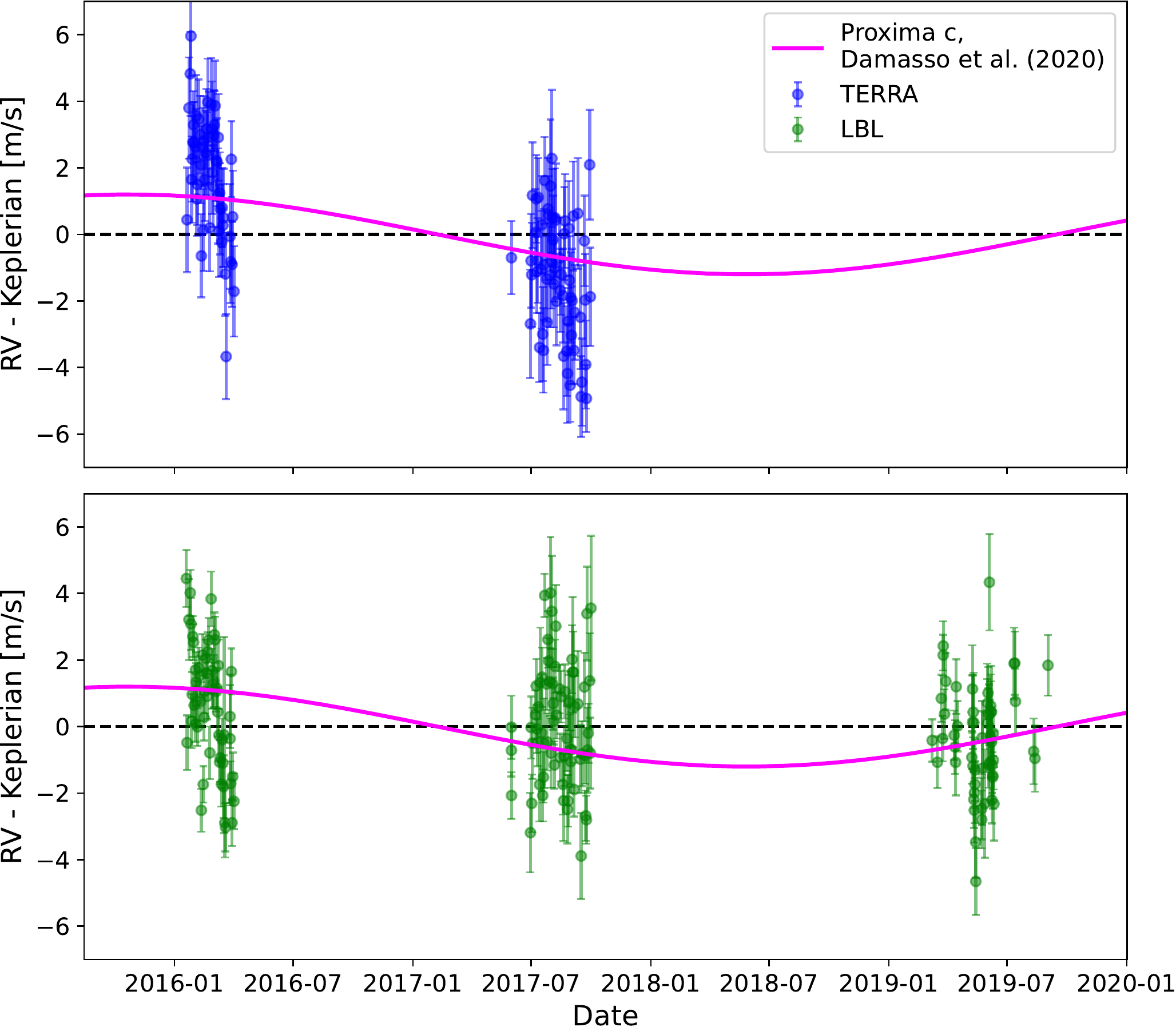}
    \caption{Residual time series for Proxima Centauri with the LBL and \texttt{HARPS-TERRA} datasets after subtracting the Proxima\,b and d Keplerian solutions of \citet{faria_candidate_2022}. The \texttt{HARPS-TERRA} time series shows a decrease of $3.2$\,m/s between the two runs, consistent with the reflex motion due to Proxima\,c. The LBL time series, extending over larger set of observations in mid-2019, shows no such trend and is inconsistent with the \cite{damasso_low-mass_2020} orbital solution for Proxima\,c.}
    \label{fig:fig_qui_tue}
\end{figure}

While pRV time series analysis requires a proper accounting of stellar activity for accurate determination of Keplerian orbits (see Section~\ref{section:lnz}), the RV measurements prior to any activity correction are worth discussing. Figure~\ref{fig:fig_qui_tue} shows the 3 sequences analysed with the LBL and the 2 with \texttt{HARPS-TERRA} compared with the Keplerian motion expected from the candidate planet, Proxima c, proposed by \citet{damasso_low-mass_2020}. One sees that the LBL, and thus \texttt{NAIRA} (see Fig.\ \ref{fig:fit2}), data are incompatible with the velocity slope expected for the Keplerian motion of Proxima c. The \texttt{HARPS-TERRA} values do display a velocity change between the 2016 and 2017 that could be interpreted as a genuine Keplerian signal, but it was shown above that this variation can be explained by a systematic effect. A more quantitative analysis is presented later that further questions the existence of Proxima\,c. 

\subsection{LBL activity indicator on Proxima} \label{sec:proxima_activity}

Activity indicators are central in the interpretation of pRV sequence as they sample the physical processes that create activity jitter that can mimic keplerian signals. The activity indicator variations are modulated at the stellar rotation period, and therefore consist of independent observations useful for detrending the RV curve from activity and extract m/s-level planetary signals. One way to do this is to first model the activity tracer (e.g., photometry, FWHM, BIS, dLW) with a Gaussian process (GP), and then use this `trained' GP to correct the RVs or by jointly fitting the RV and activity indicator. While the rotation period is often referred to with a single value, it is important to keep in mind that one expects this signal to have a time-dependent periodicity. In the case of Proxima, one expects differential rotation and secular differences on the inferred period may simply reflect latitudinal changes in the distribution of active regions with time. \citet{suarez_mascareno_magnetic_2016} and \citet{wargelin_optical_2017} point to a \hbox{$\sim$7-year} activity cycle, implying that the latitudinal distribution of active regions is likely to evolve on that timescale.  Furthermore, different methods have differing latitude-weighted sensitivity. FWMH and bisector measurements are fractionally more affected by equatorial spots than photometric measurements; one therefore does not expect a perfect correspondence between measurements taken at varying epochs and methods. Here is a summary of key literature values for the rotation period of Proxima:

\begin{itemize}
    \item \citet{benedict_photometry_1998} used HST guider photometry to measure a 83.5\,days rotation period (no uncertainty reported) between 1992 and 1997 in a bandpass centered at 583\,nm.
    \item \citet{wargelin_optical_2017} in 15-year long compilation of ASAS \citep{pojmanski_all_1997} $V$-band photometry (2001--2015) found a period ranging from 77.1\,days to 90.1\,days, pointing to an evolving lightcurve and providing a strong case for differential rotation. It is noteworthy that this analysis gives the largest span of measured rotation periods and is also the longest one, covering more than two activity cycles. This compilation supersedes \citet{kiraga_age-rotation-activity_2007} and \citet{suarez_mascareno_magnetic_2016} that used earlier releases of the same time series; the rotation periods determined by these authors are consistent with the \citet{wargelin_optical_2017} analysis (See section~2 therein).
    \item \citet{newton_new_2018} using photometry obtained from 2014 to 2018 with MEarth \citep{irwin_mearth_2008} measured a $P_{\rm rot} = 89$\,days with a typical 10\% uncertainty for their whole sample. 
    \item \citet{klein_large-scale_2020} identified a period of \hbox{$89.8\pm4$\,days} through optical spectropolarimetry in a sequence contemporaneous with our 2017 dataset.
\end{itemize}

Using the LBL framework, we can infer strong constraints on the Proxima rotation period over the timespan of the dataset by using the dLW activity indicator. This sequence is compared with the CCF FWHM from the HARPS pipeline. As for the first-order derivative measurement at the core of the LBL method, the dLW has the notable advantage over the CCF FWHM of being computed in an outlier resilient way.

We modeled the two time series with a GP using the quasi-periodic covariance kernel, a widely-used kernel for modeling variability generated by stellar rotation \citep{angus_inferring_2018}. The details of our GP prescription are given in Appendix~\ref{appendix_B} and Table~\ref{tab:activity_posterior} provides the complete list of fitted hyperparameters.

\begin{figure*}[!t]
    \centering
    \includegraphics[width=0.38\linewidth]{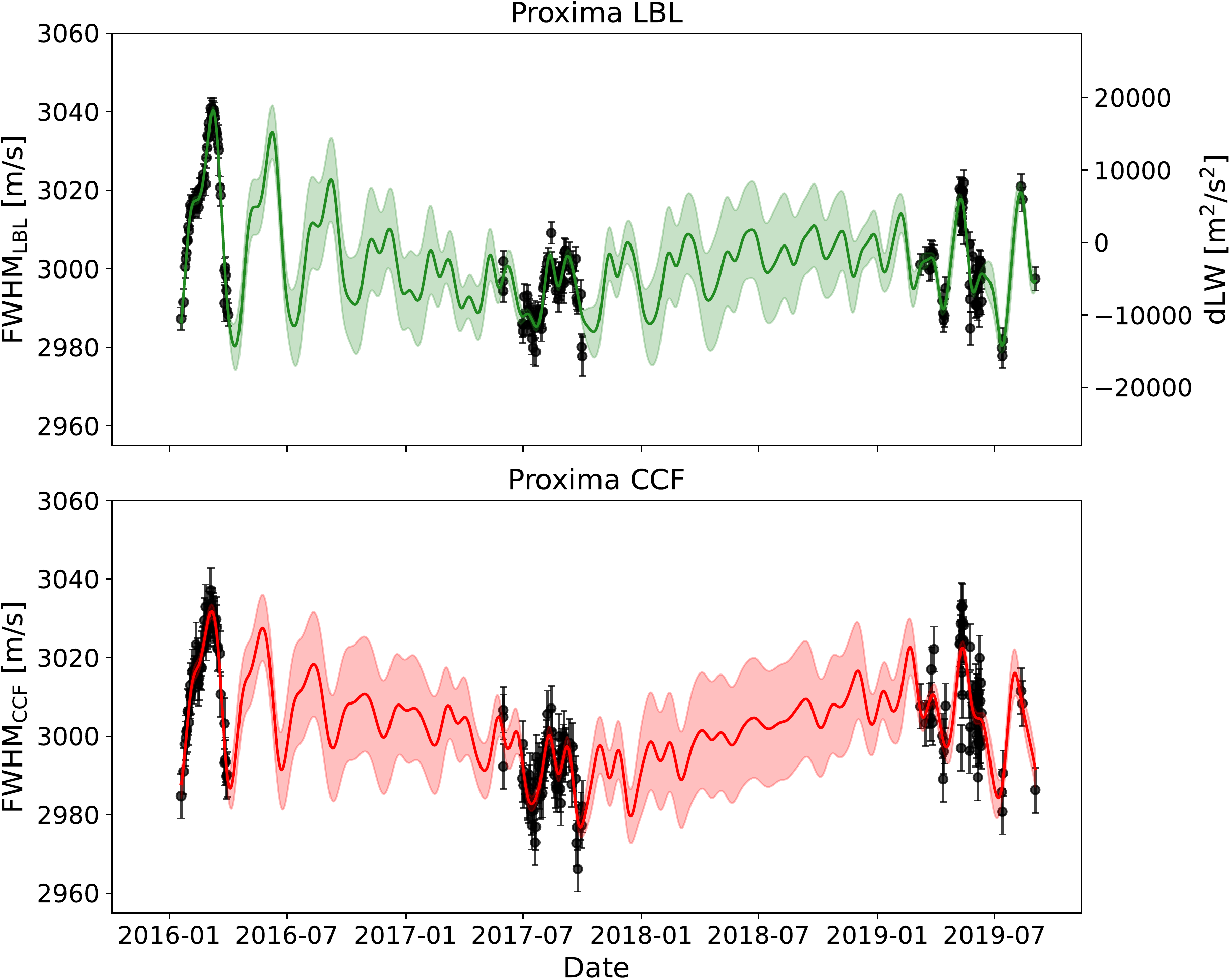}
    \includegraphics[width=0.61\linewidth]{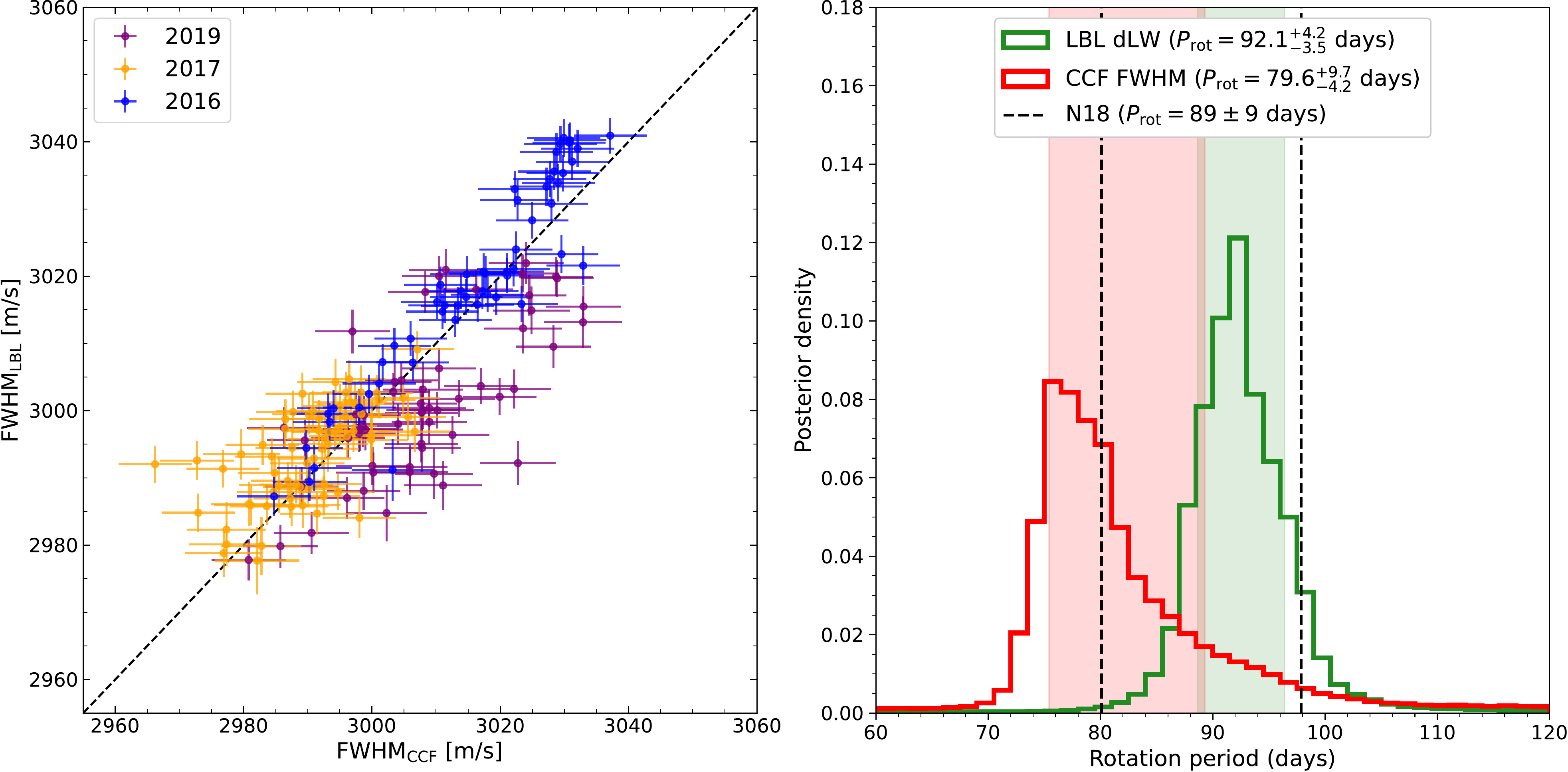}
    \caption{ Proxima Centauri activity modeling with the dLW and the CCF FWHM. \textit{Left panels}: Changes in the dLW line shape parameter compared to the CCF FWHM. The green and red curves represent the mean prediction of the best-fit quasi-periodic Gaussian process (described in Appendix~\ref{appendix_B}), each surrounded by a 68\% confidence envelope. \textit{Middle panel}: Direct comparison between the dLW (expressed in FWHM) and the CCF FWHM. One does not expect a perfect correspondence between the two variables, as the equivalence between the dLW and the FWHM is only strictly true for Gaussian lines. \textit{Right panel}: Posterior distributions on $P_{\rm rot}$ from the quasi-periodic Gaussian Process model of the dLW and CCF FWHM time series. The 68\% confidence intervals on $P_{\rm rot}$ are represented with green (LBL) and red (CCF) regions. The black dashed lines show the 10\% relative uncertainty interval for Proxima's rotation period ($P_{\rm rot} = 89$\,days) from \cite{newton_new_2018}. There is no specific uncertainty for the \cite{newton_new_2018} rotation period, but the typical uncertainties in their sample are at the 10\% level.}
    \label{fig:proxima_activity}
\end{figure*}

Figure~\ref{fig:proxima_activity} summarizes the comparison between dLW and CCF FWHM for modeling Proxima Centauri activity. The two indicators are highly correlated, particularly during the 2016 and 2019 sequences. The quasi-periodic GP models yield a rotation period of \hbox{$P_{\rm rot} = 92.1^{+4.2}_{-3.5}$\,days} for the LBL dLW and \hbox{$P_{\rm rot} = 79.6^{+9.7}_{-4.2}$\,days} for the CCF FWHM. The posterior distribution obtained with dLW is more symmetric and has a smaller 1-$\sigma$ confidence interval than the CCF-derived one. While both indicators yield periods consistent with literature values, we note that the LBL dLW provides a rotation period closer to contemporaneous measurement made by \cite{newton_new_2018} and \cite{klein_large-scale_2020}.

\subsection{Keplerian and activity model of Proxima}
\label{section:lnz}

In this section, we present a joint analysis (Keplerian and activity) of the complete Proxima LBL RV time series (2016, 2017, and 2019 sequences). We adopt the naming scheme where Proxima\,b is the 11.18-day planet \citep{anglada-escude_terrestrial_2016, suarez_mascareno_revisiting_2020, faria_candidate_2022}, Proxima\,c is the 1900-day candidate planet reported by \citet{damasso_low-mass_2020} and Proxima\,d is the 5.15\,days signal first identified as a possible companion with a $0.29\pm0.08$\,M$_{\oplus}$ minimum mass in \citet{suarez_mascareno_revisiting_2020}, later confirmed in \citet{faria_candidate_2022}. We follow this convention, acknowledging that the naming may evolve. We seek to quantify the level of each of the proposed planetary signals in radial velocities produced by the LBL algorithm.

We investigated the following suite of models ($\mathbf{M}$):
\begin{itemize}
    \item $\mathbf{M}_{\textrm{\scriptsize dLW}}$: No planet, activity GP only.
    \item $\mathbf{M}_{\textrm{\scriptsize b+dLW}}$: One planet, Proxima\,b, with activity GP.
    \item $\mathbf{M}_{\textrm{\scriptsize bd+dLW}}$: Two planets, Proxima\,b,\,d, with activity GP.
    \item $\mathbf{M}_{\textrm{\scriptsize bc+dLW}}$: Two planets, Proxima\,b,\,c, with activity GP. 
    \item $\mathbf{M}_{\textrm{\scriptsize bcd+dLW}}$: Three planets, Proxima\,b,\,c,\,d, with activity GP.
\end{itemize}

For the GP activity component, we utilized a quasi-periodic kernel trained on the dLW time series, that is with prior distributions on shared hyperparameters corresponding to the posteriors from the GP regression presented in Section~\ref{sec:proxima_activity}. For the Keplerian component of the joint fit, we allowed non-zero values for the eccentricity of Proxima\,b, while fixing $e = 0$ for Proxima\,c and Proxima\,d when included in the model. We used non-informative priors for the orbital period ($P$) and the time of inferior conjunction ($T_{0}$) of Proxima\,b (see Table~\ref{tab:posterior}). For the models with Proxima\,c, we used Gaussian priors on $P_{\rm c}$ and $T_{\rm 0,c}$ based on \cite{damasso_low-mass_2020} as our data does not cover a full period of the candidate planet. The priors on the period and phase of Proxima d were uniform, but fairly narrow around the value from \citealt{faria_candidate_2022} (see Table~\ref{tab:gp}). This is because our dataset does not include ESPRESSO data, which was mainly responsible for the detection. This model should thus be viewed as an informed search for planet d in our data, not a blind search, similarly to a transiting planet with a known period and phase. Appendix~\ref{appendix_B} provides additional details on the joint fits and Table~\ref{tab:posterior} summarizes prior and posterior distributions. 

We computed the Bayes Information Criterion (BIC, \citealt{schwarz_estimating_1978}) for each $\mathbf{M}$ considered. The smallest BIC, corresponding to the preferred model, was obtained for the single planet $\mathbf{M}_{\textrm{\scriptsize b+dLW}}$, with a BIC$_{\textrm{\scriptsize b+dLW}}$\,=\,739.1. Compared to the no planet, activity GP only solution, we measure a $\Delta$\,BIC\,$\equiv$\,BIC$_{\textrm{\scriptsize b+dLW}}-$BIC$_{\textrm{\scriptsize dLW}}$\,=\,$-37.6$, providing compelling evidence of the Proxima\,b Keplerian signal in our data. The two planets model $\mathbf{M}_{\textrm{\scriptsize bd+dLW}}$ produced a $\Delta$BIC\,=\,$-5.8$, which advocates against the necessity of adding the $\sim$0.4\,m/s Proxima\,d signal to explain our data. Taken without further input, our data and analysis would therefore be insufficient to identify this signal. This is unsurprising as our set of observations consist of a subsample of \cite{suarez_mascareno_revisiting_2020} and does not include ESPRESSO radial velocities. We nonetheless point out that the LBL yields a $P_{\rm d}$ and $K_{\rm d}$ (see Table~\ref{tab:gp} and Fig.~\ref{fig:proxima_fit}) fully consistent with the results of \cite{suarez_mascareno_revisiting_2020} and \citet{faria_candidate_2022}, with $K_{\rm d}$ detected at the 2-$\sigma$ level.

Other models including Proxima\,c are strongly rejected according to their BIC ($\Delta$BIC\,=\,$-15.1$ for $\mathbf{M}_{\textrm{\scriptsize bc+dLW}}$, $\Delta$BIC\,=\,$-21.2$ for $\mathbf{M}_{\textrm{\scriptsize bcd+dLW}}$), providing quantitative evidence against the existence of the candidate planet signal in the LBL data. This conclusion is further strengthened by the fact that the \texttt{NAIRA} analysis of the same dataset provides nearly identical results to the LBL and shows no significant RV difference between the 2016 and 2017 sequences. We note that \cite{damasso_low-mass_2020} did not clearly identify this signal in their analysis of the HARPS data only (not considering UVES observations), but the inclusion of the 2016 and 2017 sequences had a large impact on the final period and increased by 25\% the semi-amplitude (Fig.\ 2, therein). Combined with the absence of a clear detection of Proxima\,c in imaging and in astrometry \citep{gratton_searching_2020, kervella_orbital_2020, benedict_moving_2020}, our results cast serious doubts on the existence of this planet.

\section{Conclusions\label{conclusions}}
The results described here illustrate the usefulness of the LBL algorithm in handling systematic uncertainties in pRV. It outperforms, in terms of raw RV uncertainties, the CCF method as it fully utilizes the radial velocity content of the input spectra. At infrared wavelengths, the LBL algorithm opens the door to velocimetry at the m/s level as exemplified by the constraints added to the candidate planet around Barnard's star presented here and mass measurements with SPIRou of planets uncovered through TESS photometry.

At optical wavelengths, the method appears on par with template matching method in terms of accuracy, which is not overly surprising as both methods use the full RV content in spectra, but LBL appears to display lower level of systematic drifts over timescales of many months as it can efficiently reject part of the spectra affected by telluric residuals. Benchmarking over tens of targets would be necessary to disentangle the exact causes of differences in BERV trends.

A number of byproducts of the LBL algorithm could be used to further constrain activity signals. The velocity spectrum, as shown in Figure~\ref{trumpet}, provides the opportunity to measure the chromatic velocity signal in a self-consistent manner with self-consistent uncertainties. This has proved to be a useful tool in disentangling activity signals for very active objects (e.g., \citealt{zechmeister_spectrum_2018}, \citealt{cale_diving_2021}). Sorting lines by their strength may allow for the probing of different depths in the stellar photosphere and could be affected differentially by activity. This avenue has not yet been explored systematically.

As detailed in Section~\ref{sec:activity_indicator} and in \cite{zechmeister_spectrum_2018}, the projection on the second derivative is a very useful tool for activity monitoring, providing a more robust tool than the CCF FWHM for the determination of the rotation period in the Proxima Centauri data (see Figure~\ref{fig:proxima_activity}). Projections onto higher-order derivatives could be interesting as they would provide information on changes in higher moments of the line profiles. This would be conceptually analogous to the approach of \citet{collier_cameron_separating_2021} where activity is measured by a decomposition of the spectrum autocorrelation function into principal components. Numerically, we have found that the projection of the third derivative of the line profile is directly linked to the bisector tilt that traces changes in line symmetry, though we have been unable to find an analytical correspondence similar to the one presented in Section~\ref{sec:activity_indicator} linking the change in the second order derivative with changes in the FWHM. Assessing the performance of the third-order derivative as an activity indicator in comparison with indicators such as the bisector slope changes is the subject of ongoing work. Similarly to chromatic activity, a future avenue to account for stellar activity is to measure velocity with a selected subset of lines, such as done in \citep{bellotti_mitigating_2022}. This has been done in the case of Sun-like stars \citep{dumusque_measuring_2018}, but has yet to be explored for M dwarfs and in the nIR.

Our analysis of the HARPS Proxima Centauri data is a warning that  long-period low-mass planet characterized by small semi-amplitudes  at the level of a few m/s  may be resulting from unaccounted systematic effects and warrants a reanalysis of existing data using the LBL framework.

\acknowledgments

Based on observations obtained at the Canada-France-Hawaii Telescope which is operated from the summit of Maunakea by the National Research Council of Canada, the Institut National des Sciences de l'Univers of the Centre National de la Recherche Scientifique of France, and the University of Hawaii. The observations at the Canada-France-Hawaii Telescope were performed with care and respect from the summit of Maunakea which is a significant cultural and historic site. This work is party supported by the Natural Science and Engineering Research Council of Canada, the Fonds Qu\'eb\'ecois de Recherche  (Nature et Tehcnologie) the Trottier Family Foundation through the Institute for Research on Exoplanets.

The authors thank Rodrigo F. D\'iaz for suggestions regarding the use of mixture model formalism in the LBL framework.

E.A. thanks Thomas Davy for the fruitful discussions regarding the underlying logic of the pRV measurements with SPIRou that ultimately led to the current work.

J.-F.D. acknowledges funding from the European Research Council under the H2020 research \& innovation programme (grant \#740651 NewWorlds)

X.D. and A.C. acknowledge funding from ANR of France under contract number ANR\-18\-CE31\-0019 (SPlaSH).
This work is supported by the French National Research Agency in the framework of the \textit{Investissements d'Avenir} program (ANR-15-IDEX-02),  through the funding of the ``Origin of Life'' project of the \textit{Universit\'e Grenoble Alpes}.

N. A.-D. acknowledges the support of FONDECYT project 3180063.

This research made use of the following software tools:
\begin{itemize}
    \item \texttt{Astropy}; a community-developed core Python package for Astronomy. \citet{the_astropy_collaboration_astropy_2013,the_astropy_collaboration_astropy_2018}.
    \item \texttt{bottleneck}; \href{https://bottleneck.readthedocs.io}{bottleneck.readthedocs.io}.
    \item  \texttt{ds9}; a tool for data visualization supported by the Chandra X-ray Science Center (CXC) and the High Energy Astrophysics Science Archive Center (HEASARC) with support from the JWST Mission office at the Space Telescope Science Institute; \citet{joye_new_2003-1}.
    \item \texttt{emcee}; \citet{foreman-mackey_emcee_2013}.
    \item \texttt{george}; \citet{ambikasaran_fast_2016}.
    \item \texttt{IPython}; \citet{perez_ipython_2007}.
    \item \texttt{matplotlib}, a Python library for publication quality graphics; \citet{hunter_matplotlib_2007}. 
    \item \texttt{Numba}; \citet{lam_numba_2015}.
    \item \texttt{NumPy}; \citet{harris_array_2020}.
    \item \texttt{radvel}; \citet{fulton_radvel_2018}.
    \item \texttt{SciPy}; \citet{scipy_10_contributors_scipy_2020}.
    \item \texttt{tqdm}; \citet{da_costa-luis_tqdm_2021}.
\end{itemize}

\facilities{CFHT(SPIRou), ESO La Silla (HARPS),  ESO Science Archive.}

\clearpage
\appendix
\counterwithin{table}{section}
\counterwithin{figure}{section}

\section{Code accessibility}
\label{code_accessibility}
The most up-to-date implementation of the LBL algorithm is available to the community at \href{https://lbl.exoplanets.ca/}{lbl.exoplanets.ca}, with a link to the relevant GitHub repository and sample datasets. Tools to generate the templates and masks are also provided. The \textit{rdb} format output of the LBL codes is compatible with the {\it Data \& Analysis Center for Exoplanets\footnote{\href{https://dace.unige.ch/dashboard/}{dace.unige.ch/dashboard/}}} online tools to analyse pRV time series.

\section{Weighted mean in the presence of outliers}\label{appendix_A}

The LBL method requires the combination of a large number of velocities for which uncertainty estimates are known, but allowing for the possibility that a small fraction of values are drawn from a population of spurious values for which we have no prior knowledge of the expected distribution. The fraction of lines expected to belong to the spurious population can be estimated from the fraction of high-$\sigma$ outliers in the velocity distribution. Assuming that the outlier represents a fraction $f_0$ of all lines, that each $i$ point has an uncertainty estimate $\sigma_i$, a value $v_i$, a weighted mean velocity $\overline{v}$, the normalized distance to mean $\sigma_\mu = \frac{v_i - \overline{v}}{\sigma_i}$, the likelihood that one point is a valid point is:
\[
p_i = \frac{(1-f_0) e^{-\sigma_\mu^2/2}}{(1-f_0) e^{-\sigma_\mu^2/2}+f_0 }
\]

The mean velocity is assumed to be the same for all lines and can be determined through a weighted sum that takes into account both the likelihood of the point being valid and the uncertainties as derived from the LBL algorithm:
\[
\overline{v} = \frac{\sum_{i} v_i  p_i \sigma_i^{-2} }{\sum_{i} p_i \sigma_i^{-2}}
\]
The corresponding uncertainty for the mean velocity is:
\[
\overline{\sigma} = \sqrt{\frac{1 }{\sum_{i} p_i \sigma_i^{-2}}}
\]

The above equations are coupled, but converge rapidly with an iterating algorithm. One starts with an approximation of $\overline{v}$ by using the median value of $v_i$ to determined $p_i$ and use these value to refine the $\overline{v}$ value. Assuming that $f_0$ is very small (typically $<0.1\%$ in SPIRou data), $\overline{v}$ converges to a small fraction of $\overline{\sigma}$ within two iterations. We compiled the number of outliers in the Barnard's star data, consisting of $4.6\times10^6$ individual line measurements (247 spectra and $\sim19\,000$ lines per spectrum) and found that $5.8\times10^{-4}$ of the lines are $>5$-$\sigma$ outliers. This is a much larger fraction than would be expected from a normal distribution ($\sim6\times10^{-7}$) and suggests that these lines belong in vast majority to the 'outlier' population.

\begin{figure*}[!htbp]
    \centering
    \includegraphics[width=0.45\linewidth]{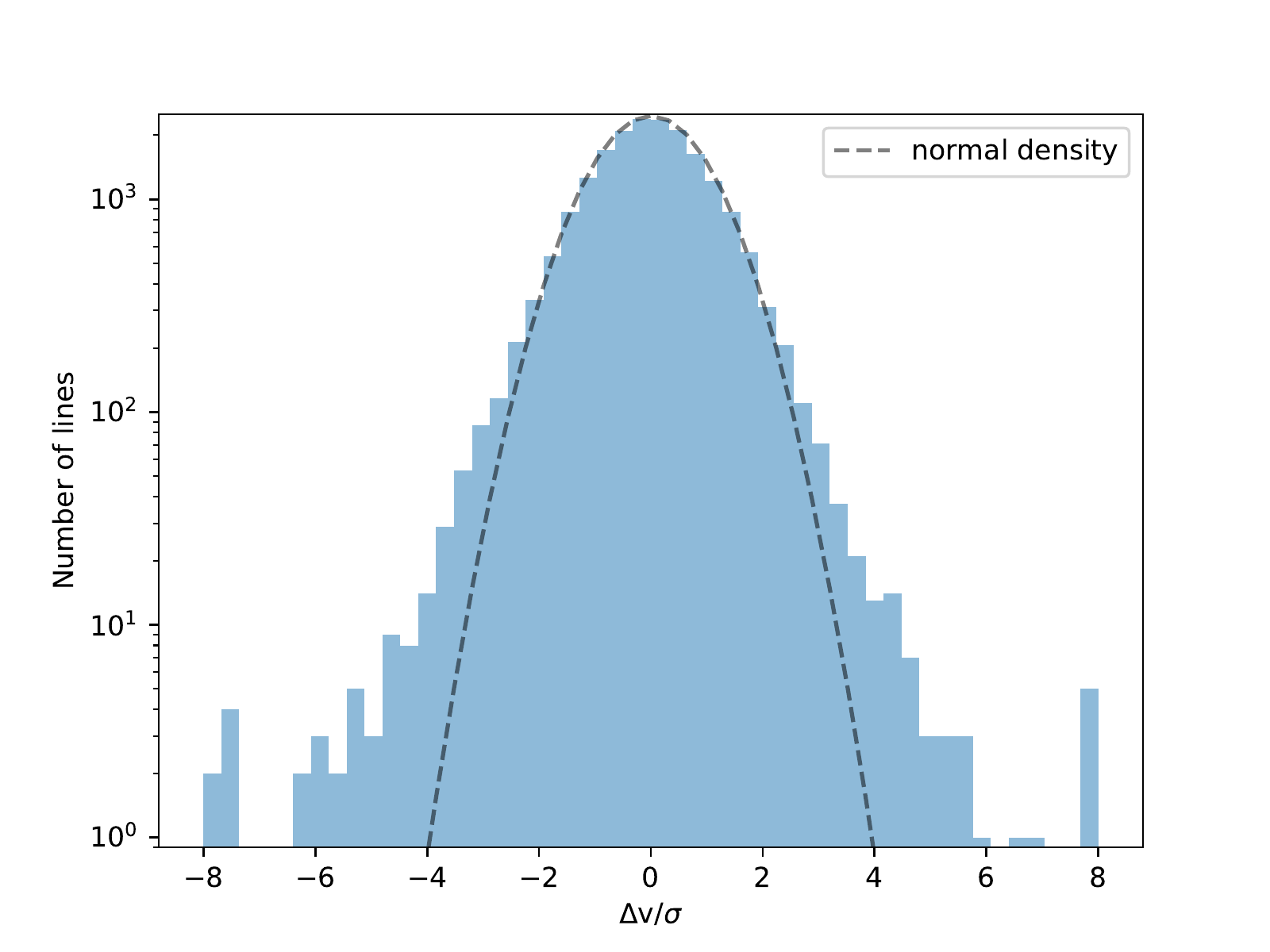}
    \includegraphics[width=0.45\linewidth]{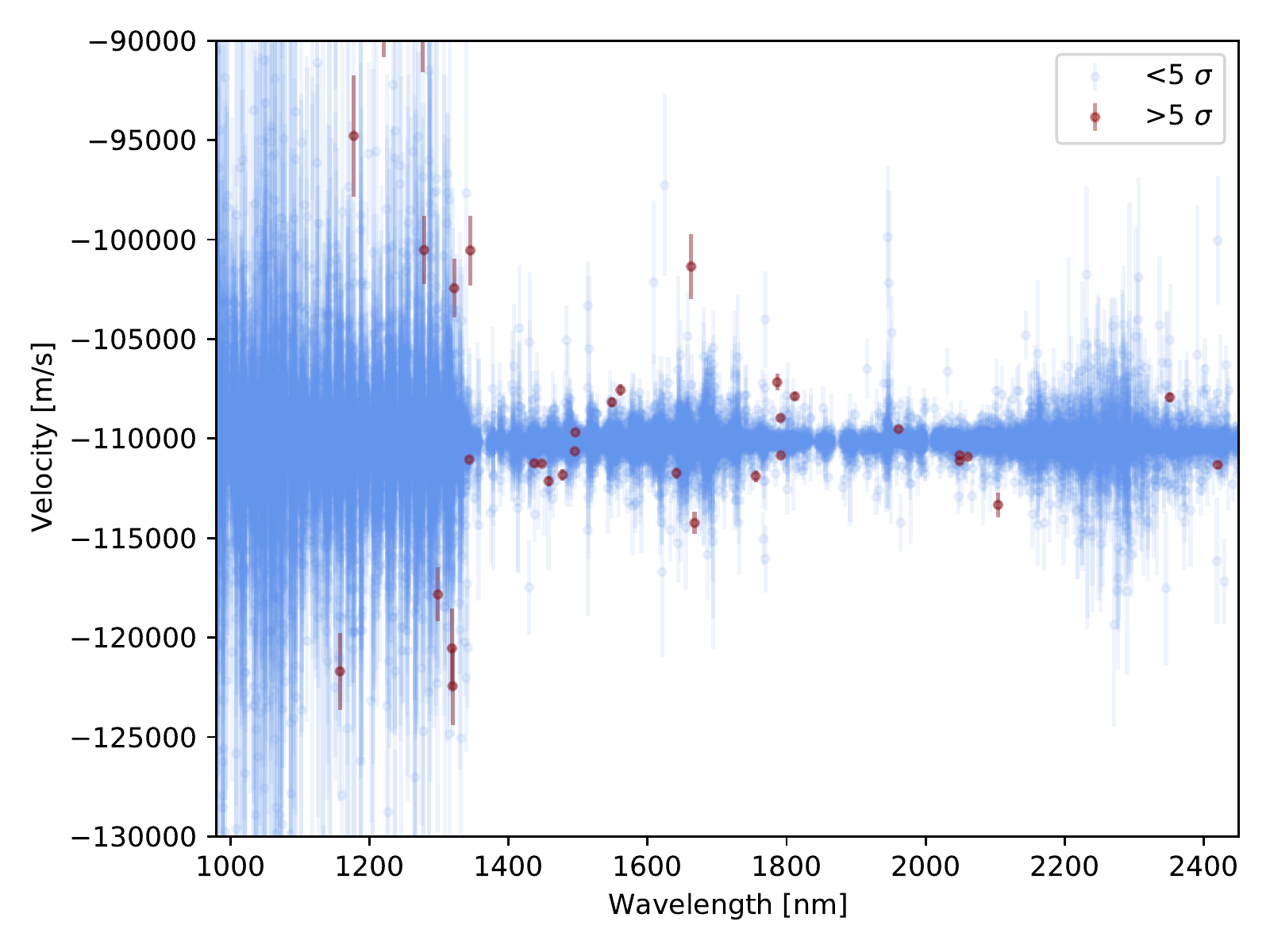}

    \caption{\textit{Left}: For a sample Barnard's star observation, the histogram of the per-line velocity difference from the mean velocity divided by the corresponding line uncertainty closely follows a normal distribution (dashed line) with only a handful of $>5\sigma$ outliers. \textit{Right}: Spectral distribution of $>5\sigma$ outliers in a random Barnard's star observations.
    }
    \label{fig:sigma_dist}
\end{figure*}

To determine the experimental $f_0$ value, we performed a Gaussian fit on the density distribution of $\frac{v_i-\overline{v}}{\sigma}$ (see Figure~\ref{fig:sigma_dist}), subtracted the fit from the distribution and found an excess of $2\times10^{-3}$ of values beyond $\pm3\sigma$ compared to a normal distribution. We therefore adopt a $f_0 = 2\times10^{-3}$ for all our analysis but  the resulting pRV time series are largely insensitive to changes in $f_0$ ranging from $10^{-2}$ to $10^{-4}$. An $f_0=2\times10^{-3}$ will lead to a $p_i=0.5$ for a $3.5\sigma$ event and $p_i\sim0.02$ for a $4.5\sigma$ event. In practice, this summation scheme is equivalent to a $\sim$4-$\sigma$ clipping with soft edges rather than binary rejection. This weighted mean is used both for the determination of the mean velocity and dLW with the same $f_0$ value.

\section{Material regarding GP models}\label{appendix_B}

In Section~\ref{sec:proxima_activity}, we modeled Proxima stellar activity in the LBL dLW and the HARPS pipeline CCF FWHM time series with a Gaussian Process. We employed the following quasi-periodic covariance kernel:
\begin{equation}
k_{i,j} = A^2 \exp \left[ -\frac{|t_i - t_j|^2}{2 \ell^2} - \Gamma^2 \sin^2 \left( \frac{\pi | t_i - t_j|}{P_{\rm rot}} \right) \right] + s^2 \delta_{i,j}
\end{equation}

where $|t_i - t_j|$ is the time interval between data $i$ and $j$, $A$ is the amplitude of the GP, $\ell$ is the correlation timescale, $\Gamma$ scales the periodic component of the GP, $P_{\rm rot}$ is the stellar rotation period, and $s$ is an extra white noise term added in the diagonal of $k_{i,j}$. Our GP modelization consisted of 5 hyperparameters all fitted in logarithmic form \{$\ln A$, $\ln \ell$, $\ln \Gamma$, $\ln P_{\rm rot}$, $\ln s$\}. We used \texttt{george} \citep{ambikasaran_fast_2016} to compute GP predictions and calculate the log-likelihood. The parameter posterior distributions were sampled using a MCMC algorithm with \texttt{emcee} \citep{foreman-mackey_emcee_2013}. We used 100 walkers, ran 25\,000 steps, and rejected (burn-in) the first 2000 samples. The number of steps was greater than 50 times the autocorrelation timescale (computed by \texttt{emcee}), which is generally an indication of convergence (\citealt{dewitt-morette_monte_1997}; \citealt{foreman-mackey_emcee_2019}). Table~\ref{tab:activity_posterior} lists the prior distributions adopted for the hyperparameters of the dLW and CCF FWHM GP regressions, as well as their posterior's median, 16$^{\rm th}$ and 84$^{\rm th}$ percentiles. Figure~\ref{fig:proxima_activity} shows the mean GP predictions for the dLW and the CCF FWHM, using the best-fit hyperparameters.

\begin{figure*}[!htbp]
    \centering
    \includegraphics[width=0.45\linewidth]{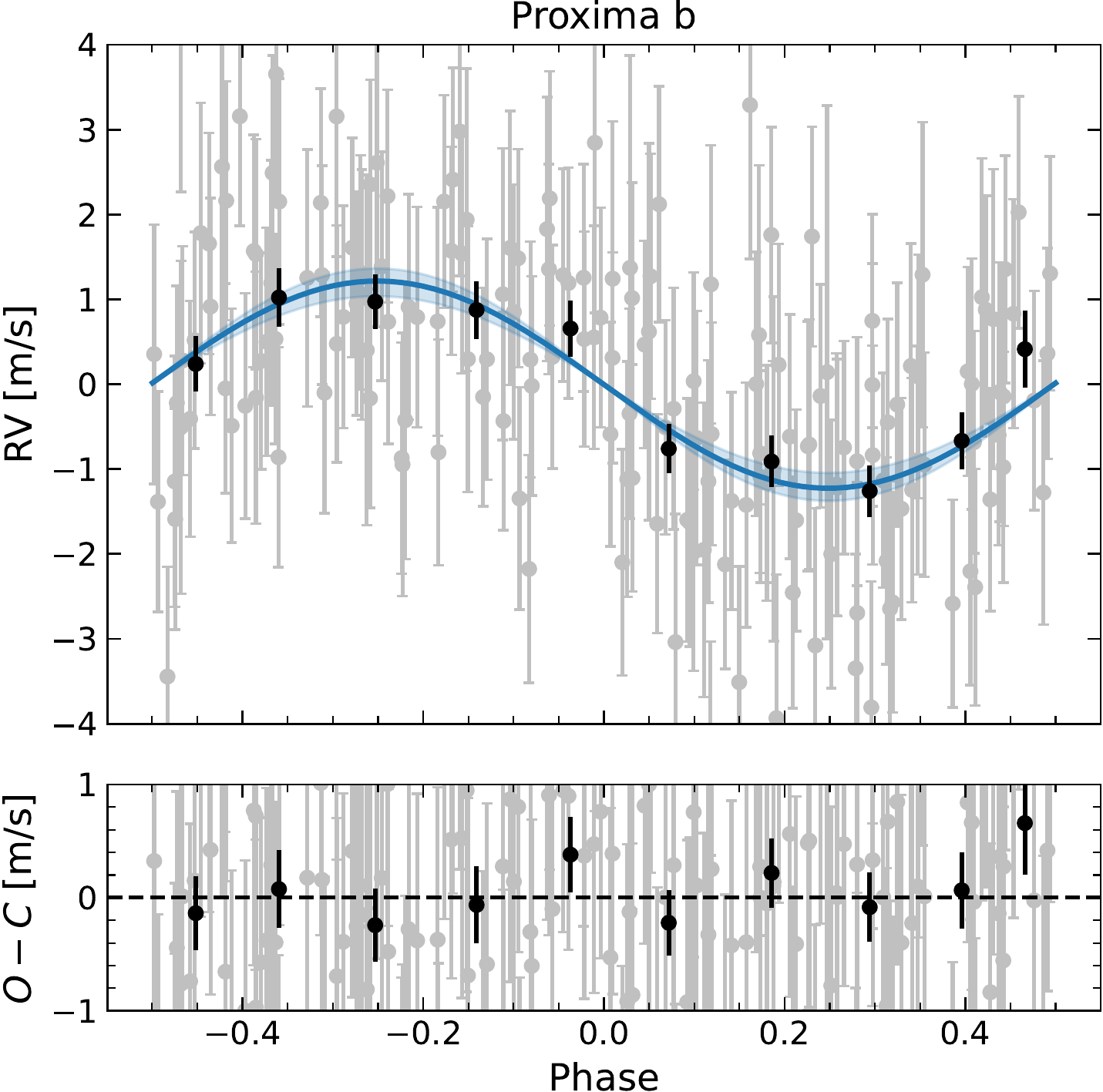}
    \includegraphics[width=0.45\linewidth]{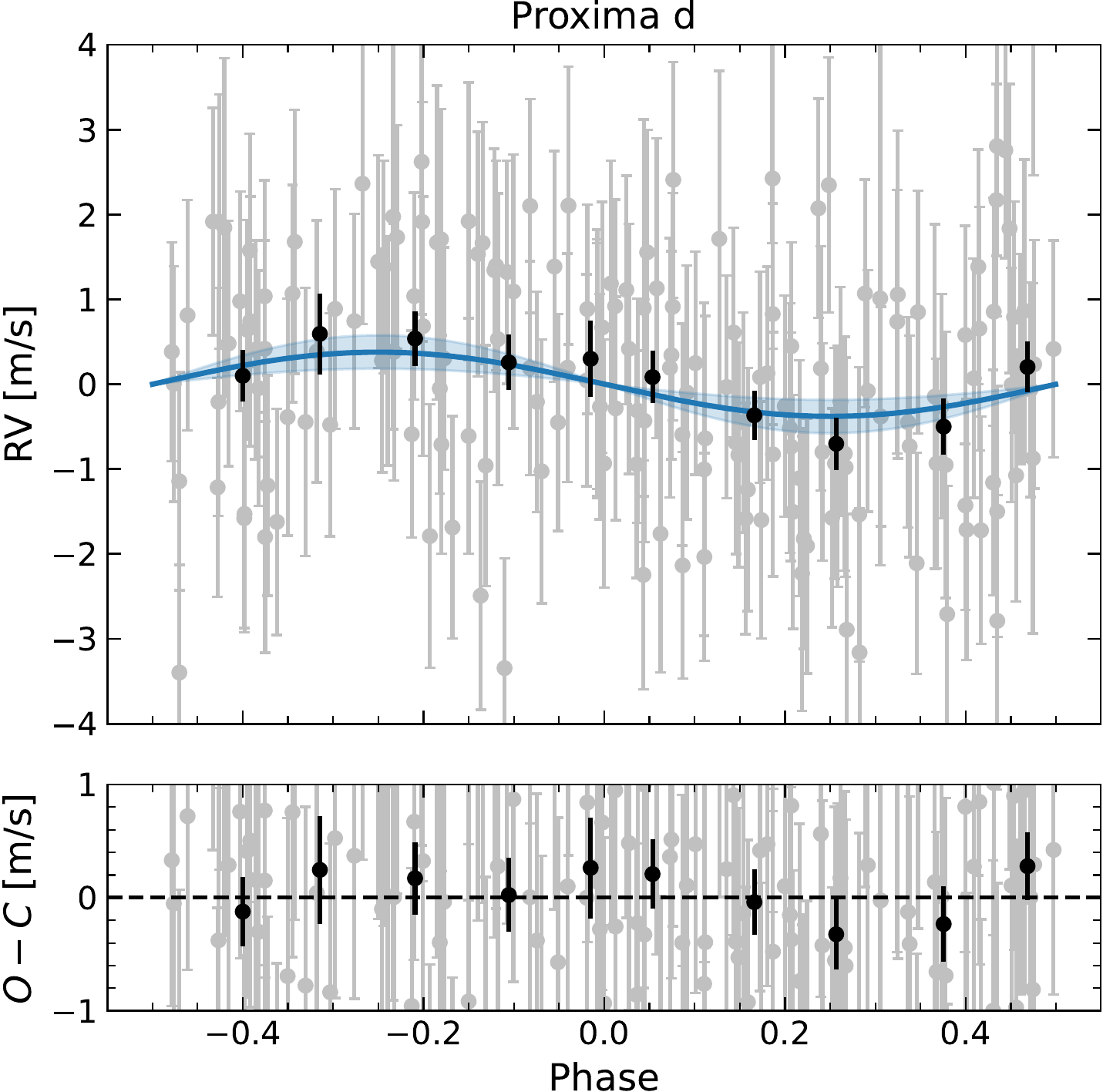}
    \caption{Phase-folded RV curves (top panels) and orbital fit residuals (bottom panels) for Proxima b and d (model $\mathbf{M}_{\textrm{\scriptsize bd+dLW}}$), with systemic velocity and activity GP removed. Binned RVs (0.1 phase bin) are represented with black data points. The best-fit Keplerian models for Proxima\,b and d are shown with solid blue curves, with 68\% confidence interval envelopes in light blue.}
    \label{fig:proxima_fit}
\end{figure*}

In Section \ref{section:lnz}, we jointly fit Keplerian signals and activity in the LBL radial velocities for 5 different models ($\mathbf{M}_{\textrm{\scriptsize dLW}}$, $\mathbf{M}_{\textrm{\scriptsize b+dLW}}$, $\mathbf{M}_{\textrm{\scriptsize bd+dLW}}$, $\mathbf{M}_{\textrm{\scriptsize bc+dLW}}$, and $\mathbf{M}_{\textrm{\scriptsize bcd+dLW}}$). We generated Keplerian orbits with \texttt{radvel} \citep{fulton_radvel_2018} and used the same quasi-periodic kernel as described above for the GP activity component. We adopted as prior distributions on the hyperparameters $\ln \ell$, $\ln \Gamma$, and $\ln P_{\rm rot}$, the posteriors of the GP regression of the dLW time series (Table~\ref{tab:activity_posterior}). This is justified because stellar activity processes affecting the RVs and the dLW are expected to share common properties (correlation timescale, periodicity, etc.) described by the hyperparameters. We sampled the posterior distributions for each $\mathbf{M}$ with \texttt{emcee}, using 100 walkers, 50\,000 steps, and a burn-in of 5\,000. The model favoured according to BIC comparison is the single planet $\mathbf{M}_{\textrm{\scriptsize b+dLW}}$. We describe the free parameters of model $\mathbf{M}_{\textrm{\scriptsize b+dLW}}$ and report the prior and posterior distributions in Table~\ref{tab:posterior}. The $\mathbf{M}_{\textrm{\scriptsize bd+dLW}}$ model informed on the mass determination significance of Proxima\,d in the LBL data. In Table~\ref{tab:gp}, we compare our fitted values of $P$ and $K$ for Proxima\,b and d with the ones obtained in \cite{suarez_mascareno_revisiting_2020} and \citet{faria_candidate_2022}. The phase-folded best-fit Keplerian components for planet b and d are presented in Figure~\ref{fig:proxima_fit}.

The results obtained with the preferred model $\mathbf{M}_{\textrm{\scriptsize b+dLW}}$ are compatible with literature values. We measure a semi-amplitude for Proxima\,b ($K_{\rm b} = 1.27\pm0.17$\,m/s) that is fully consistent with \citealt{damasso_low-mass_2020} ($K_{\rm b} = 1.2\pm0.1$\,m/s), \citealt{suarez_mascareno_revisiting_2020} ($K_{\rm b} = 1.38\pm0.10$\,m/s) and \citealt{faria_candidate_2022} ($K_{\rm b} = 1.24\pm0.07$\,m/s). The increased uncertainty on our $K_{\rm b}$ is explained by the fact that the LBL time series consists of a sub-sample of \cite{damasso_low-mass_2020} and \cite{suarez_mascareno_revisiting_2020} dataset, focusing on post-fiber change  observations with HARPS.

\begin{deluxetable*}{c|cc|cc}
\tablecaption{Fitted hyperparameters of the dLW and the CCF FWHM Gaussian Process regressions.\label{tab:activity_posterior}}
\tablehead{
\colhead{Parameter} & \colhead{Prior} & \colhead{LBL dLW} & \colhead{Prior} & \colhead{CCF FWHM}
}
\startdata
$\ln A$ & $\mathcal{U}(-5, 15)$ & $9.0\pm0.2$ & $\mathcal{U}(-5, 5)$ & $2.7^{+0.3}_{-0.2}$\\
$\ln \ell$  & $\mathcal{U}(3, 10)$ & $5.3^{+0.4}_{-0.5}$ &  $\mathcal{U}(3, 10)$ & $4.9^{+0.5}_{-0.7}$ \\
$\ln \Gamma$  & $\mathcal{U}(-5, 5)$ & $0.43^{+0.16}_{-0.17}$ & $\mathcal{U}(-5, 5)$ & $0.27^{+0.23}_{-0.24}$ \\
$\ln P_{\rm rot}$ & $\mathcal{U}(\ln 60, \ln 120)$ & $4.523^{+0.045}_{-0.039}$ & $\mathcal{U}(\ln 60, \ln 120)$ & $4.38^{+0.11}_{-0.05}$ \\
$\ln s$ & $\mathcal{U}(-5, 15)$ & $7.15\pm0.11$ & $\mathcal{U}(-5, 5)$ & $1.73\pm0.06$ \\
\enddata
\tablecomments{The LBL dLW has m$^2$/s$^2$ units, compared to m/s for the CCF FWHM. The adopted priors for $\ln A$ and $\ln s$ differ for both indicators because of the different units. For the other parameters, $\ell$ is in days, $\Gamma$ is dimensionless, and $P_{\rm rot}$ is in days for both datasets.}
\end{deluxetable*}

\begin{deluxetable*}{lccr}
\tablecaption{Fitted orbital parameters for Proxima\,b and activity GP (model $\mathbf{M}_{\textrm{\scriptsize b+dLW}}$).\label{tab:posterior}}
\tablehead{
\colhead{Parameter} & \colhead{Prior$^1$} & \colhead{Posterior} & \colhead{Description}
}
\startdata
\multicolumn{4}{c}{\textit{Fitted parameters}}\\
$\gamma$ [m/s]  & $\mathcal{U}(-22002, -21982)$ & $-21992.2\pm0.5$ & Systemic velocity component \\
$T_{\rm 0,b}$ [BJD - 2450000]  & $\mathcal{U}(7895, 7910)$ &$7898.5\pm0.4$ & Time of inferior conjunction \\
$P_{\rm b}$ [days] & $\mathcal{U}(\mathbf{10.0}, \mathbf{20.0})$ & $11.1888^{+0.0060}_{-0.0059}$ & Orbital period \\
$K_{\rm b}$ [m/s] & $\mathcal{U}(0, 5)$ & $1.27\pm0.17$ & RV semi-amplitude \\
$h_{\rm b} = \sqrt{e_{\rm b}} \cos \omega_{\rm b}$ & $\mathcal{U}(-1, 1)$ & $-0.06^{+0.25}_{-0.23}$ &  Parametrization of $e$ and $\omega$ \\
$k_{\rm b} = \sqrt{e_{\rm b}} \sin \omega_{\rm b}$ & $\mathcal{U}(-1, 1)$ & $0.08^{+0.29}_{-0.28}$ & Parametrization of $e$ and $\omega$ \\
\hline
\multicolumn{4}{c}{\textit{Derived parameters}}\\
$M_{\rm p,b} \sin i$ [$M_{\oplus}$]\,$^2$ & -- & $1.07\pm0.14$ &  Minimal mass\\
$e_{\rm b}$ & -- & $0.10^{+0.13}_{-0.07}$ &  Orbital eccentricity\\
$\omega_{\rm b}$ [rad] & -- & $0.8^{+1.5}_{-2.7}$ & Argument of periastron \\
\hline
\multicolumn{4}{c}{\textit{Activity GP parameters}}\\
$\ln A$  & $\mathcal{U}(-5, 5)$ & $0.35^{+0.25}_{-0.24}$ & Amplitude of the GP \\
$\ln \ell$  & $\mathcal{GP}(\textrm{dLW})$ & $5.3^{+0.4}_{-0.5}$ &  Coherence length of the GP \\
$\ln \Gamma$  & $\mathcal{GP}(\textrm{dLW})$ & $0.52\pm0.14$ & Periodic component scale  \\
$\ln P_{\rm rot}$  & $\mathcal{GP}(\textrm{dLW})$ & $4.515^{+0.032}_{-0.037}$ & Stellar rotation period  \\
$\ln s$  & $\mathcal{U}(-5, 5)$ & $0.12\pm0.09$ & Extra white noise term \\
\enddata
\tablecomments{ $^1\mathcal{GP}(\textrm{dLW})$ is the posterior distribution on the hyperparameter from the dLW GP regression. $^2$Using the \cite{mann_how_2015} estimate of Proxima mass ($M_{\star} = 0.1221\pm0.0022$\,M$_{\odot}$) calculated in \cite{suarez_mascareno_revisiting_2020}.}
\end{deluxetable*}

\begin{deluxetable*}{ccccc}
\tablecaption{Comparison between the LBL detection of Proxima\,b and d (model $\mathbf{M}_{\textrm{\scriptsize bd+dLW}}$) with published literature values.}
\tablehead{
\colhead{Parameter} & \colhead{Prior} & \colhead{$^1$LBL} & \colhead{$^2$SM2020} & \colhead{$^3$Fa2022}
}
\startdata
$P_{\rm b}$ [days] & $\mathcal{U}(\mathbf{10.0}, \mathbf{20.0})$ & $11.1881^{+0.0061}_{-0.0058}$ & $11.1842\pm0.0007$ & $11.1868^{+0.0029}_{-0.0031}$\\[0.2cm]
$K_{\rm b}$ [m/s] & $\mathcal{U}(\mathbf{0}, \mathbf{5})$ & $1.22\pm0.17$ & $1.37\pm0.10$ & $1.24\pm0.07$\\
\hline
$P_{\rm d}$ [days] & $\mathcal{U}(\mathbf{5.0}, \mathbf{5.3})$ & $5.167^{+0.047}_{-0.091}$ & $5.168^{+0.051}_{-0.069}$ & $5.122^{+0.002}_{-0.036}$ \\[0.2cm]
$K_{\rm d}$ [m/s] & $\mathcal{U}(\mathbf{0}, \mathbf{5})$ & $0.38^{+0.19}_{-0.20}$ & $0.35^{+0.10}_{-0.11}$ & $0.39\pm0.07$\\
\enddata
\tablecomments{$^1$This work, from publicly available HARPS data (post-fiber change, $N = 181$).\\
$^2$\cite{suarez_mascareno_revisiting_2020} values, Table\,3 (GP+2 Signals) therein, from UVES ($N_{\rm UVES} = 63$), HARPS ($N_{\rm HARPS} = 196$), and ESPRESSO ($N_{\rm ESPRESSO} = 77$) data.\\
$^3$\citet{faria_candidate_2022} values, from their Table\,C.1 using the template matching RVs from ESPRESSO ($N = 114$).}
\label{tab:gp}
\end{deluxetable*}

\bibliography{lbl.bib}{}
\bibliographystyle{aasjournal.bst}
\end{document}